\documentclass{jfm}

\usepackage{graphicx}
\usepackage{newtxtext}
\usepackage{newtxmath}
\usepackage{natbib}
\usepackage{hyperref}
\hypersetup{
    colorlinks = true,
    urlcolor   = blue,
    citecolor  = black,
}

\newcommand{\RomanNumeralCaps}[1]
\linenumbers

\usepackage{mathtools}
\usepackage{amsmath}
\usepackage{amssymb}

\title{On the role of vorticity stretching and strain self-amplification in the turbulence energy cascade}

\author{Perry L. Johnson
  \corresp{\email{perry.johnson@uci.edu}}}

\affiliation{Department of Mechanical and Aerospace Engineering, University of California, Irvine, CA, USA}

\begin{document}
\maketitle

\begin{abstract}
The tendency of turbulent flows to produce fine-scale motions from large-scale energy injection is often viewed as a scale-wise cascade of kinetic energy driven by vorticity stretching. This has been recently evaluated by an exact, spatially-local relationship [Johnson, P. L. 2020, Phys. Rev. Lett., \textbf{124}, 104501], which also highlights the contribution of strain self-amplification. In this paper, the role of these two mechanisms is explored in more detail. Vorticity stretching and strain amplification interactions between velocity gradients filtered at the same scale account for about half of the energy cascade rate, directly connecting restricted Euler dynamics to the energy cascade. Multiscale strain amplification and vorticity stretching are equally important, however, and more closely resemble eddy viscosity physics. Moreover, ensuing evidence of a power-law decay of energy transfer contributions from disparate scales supports the notion of an energy cascade, albeit a `leaky' one. Besides vorticity stretching and strain self-amplification, a third mechanism of energy transfer is introduced and related to the vortex thinning mechanism important for the inverse cascade in two dimensions. Simulation results indicate this mechanism also provides a net source of backscatter in three-dimensional turbulence, in the range of scales associated with the bottleneck effect. Taken together, these results provide a rich set of implications for large-eddy simulation modeling.
\end{abstract}



\section{Introduction}
\label{sec:intro}

One of the most salient features of turbulent flows is enhanced rates of energy dissipation ($\epsilon$) and mixing. Turbulence rapidly produces fine-scale variations in the velocity field. The dynamics of this process has been of enduring interest for understanding and modeling turbulent flows across a wide range of applications. The resulting broadband range of length and time scales, evolving under intrinsically nonlinear dynamics, lies at the center of what makes turbulent flows challenging to analyze, model, and predict.

The kinetic energy cascade is the predominant concept for illuminating the production of small scales in turbulence \citep{Richardson1922, Kolmogorov1941a, Onsager1949, Frisch1995}. Turbulent kinetic energy is produced primarily in the form of large-scale motions with size comparable to that of the shear flow that serves as the energy source. The kinetic energy is dissipated to thermal energy by viscosity ($\nu$) primarily acting on the smallest-scale motions in the flow, which are comparable in size to the Kolmogorov length scale, $\eta = \nu^{3/4} \epsilon^{-1/4}$ \citep{Kolmogorov1941a}. The energy cascade describes the process of transferring kinetic energy from the largest-scale motions where it is produced to the smallest-scale motions responsible for irreversible dissipation. The cascade phenomenology asserts that the predominant exchanges of energy occur between coherent motions having nearly the same size, so that energy is passed in a quasi step-wise manner to successively smaller-scale motions. The cascade terminates when the smallest scales are finally energized and viscous dissipation removes kinetic energy at the rate it is supplied.

The energy cascade provides a conceptually simple phenomenological explanation for the observed complex, chaotic behavior of turbulence. Supposing that the energy transfer processes are chaotic and somewhat independent across length scales, the cascade idea provides an attractive explanation for why universal properties are observed for such a wide range of turbulent flow scenarios. However, two significant questions are raised.

First, is the net transfer of energy from large to small scales in turbulence actually accomplished primarily by interactions between motions having nearly the same size? This is the question of scale-locality, which is a necessary property for the cascade phenomenology to be plausible. Strictly speaking, the energy transfer does not occur in a well-defined step-like manner \citep{Lumley1992}. In this sense, the cascade is `leaky.' However,  the question of scale-locality may be answered more rigorously by quantifying the relative contribution of interactions between differently-sized motions to the energy cascade rate. Theoretical analysis predicts that the contribution to the cascade rate across scale $\ell$ due to interactions with motions at scale $\ell^\prime < \ell$ decays as a power-law for $\ell^\prime \ll \ell$, specifically, $\sim \left( \ell^\prime / \ell \right)^{4/3}$ \citep{Kraichnan1966, Kraichnan1971, Eyink2005, Eyink2009}. This power-law indicates the degree of ultra-violet locality \citep{Lvov1992}. Numerous numerical investigations have revealed at least some degree of scale-locality in the mean energy transfer across scales, with general support for the theoretical scaling \citep{Zhou1993a, Zhou1993b, Aoyama2005, Mininni2006, Domaradski2007a, Domaradski2007b, Mininni2008, Eyink2009, Aluie2009, Domaradski2009, Cardesa2015, Doan2018}.

\begin{figure}
	\centering
	\includegraphics[width=1.0\linewidth]{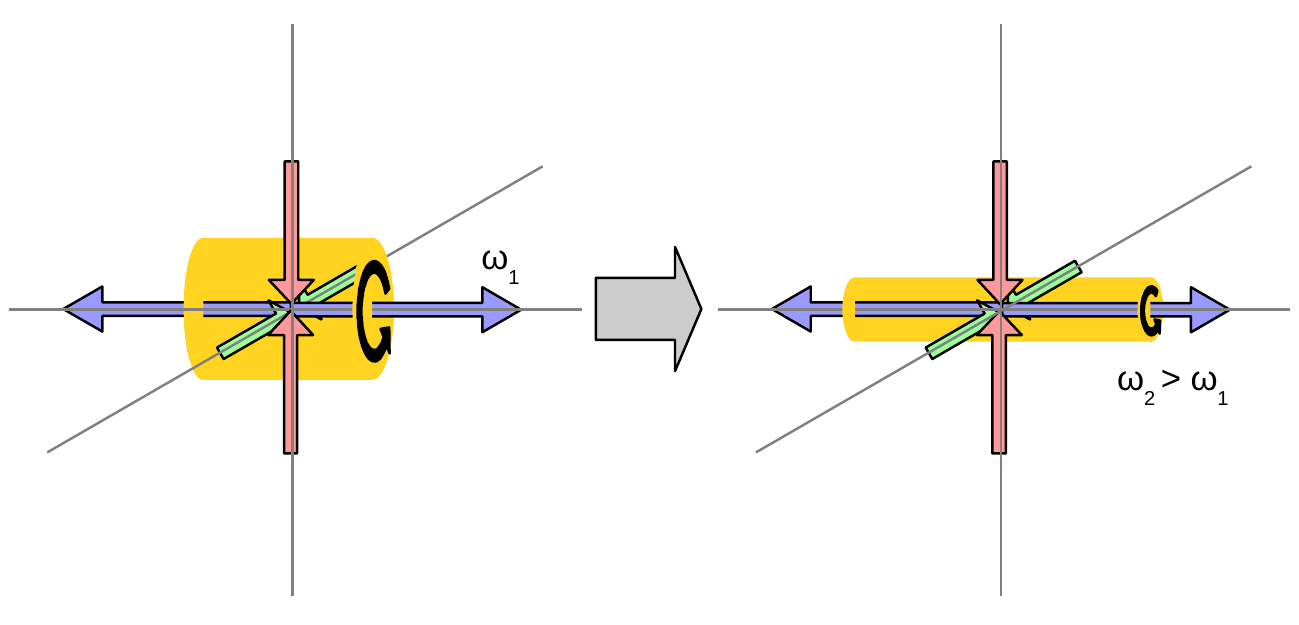}
	\caption{A simplified illustration of vortex stretching.}
	\label{fig:vortex-stretching}
\end{figure}

The second question of significant importance to the energy cascade phenomenology is: what dynamical mechanism is responsible for driving such a cascade? To some degree, answering this second question should also shed some light on the first. That is, identifying the dynamical mechanism of the cascade should provide some basis for examining the scale-locality of that mechanism. For example, consider the most commonly cited mechanism for the cascade, i.e., vortex stretching \citep{Taylor1938, Onsager1949, Tennekes1972}. The vortex stretching mechanism of the energy cascade is illustrated in Figure \ref{fig:vortex-stretching} and may be described as follows. A compact region characterized by elevated vorticity (i.e., a vortex) subjected to an extensional strain-rate along its axis of rotation will have its cross-section reduced. As this occurs, the vorticity in that region will be enhanced in keeping with Kelvin's theorem. In this way, kinetic energy is passed from straining motions to vortical motions that decrease in size. Owing to this proposed cascade mechanism, the role of vortex structures in turbulence dynamics has been an active subject of research over the years \citep{Lundgren1982,Chorin1988, Pullin1994, Pullin1998, Jimenez1998, Lozano2016}. For instance, numerical simulations have been used to confirm that interactions between vortices and straining regions are strongest when the vortical structure is only somewhat smaller than the straining region \citep{Doan2018}.

The question of dynamical mechanism is of high importance in its own right \citep{Carbone2020}. This is especially true for reduced-order models of turbulence, a practical step for predicting many turbulent flows. Large eddy simulations (LES) are widely-used and increasing in popularity with growing computational resources. The LES approach directly calculates the dynamics of large-scale motions on a computational grid while requiring a closure model for representing the effect of smaller, unresolved scales on the computed large-scale motions. Arguably the most important feature of an LES closure model is that it remove energy from the resolved scales in an accurate manner. However, eddy viscosity closures remain the most common approach to LES alongside heuristic approaches which rely on implicit modeling via specially-designed discretization errors. From a physical perspective, these popular models are known to be deficient despite their widespread use \citep{Borue1998}. In contrast, the stretched vortex model of \citet{Misra1997} attempts to build a model on more explicit physical grounds, appealing to the classical vortex stretching mechanism of the cascade. It has also been proposed to use vortex stretching to determine an eddy viscosity \citep{Silvis2019}.

\begin{figure}
	\centering
	\includegraphics[width=1.0\linewidth]{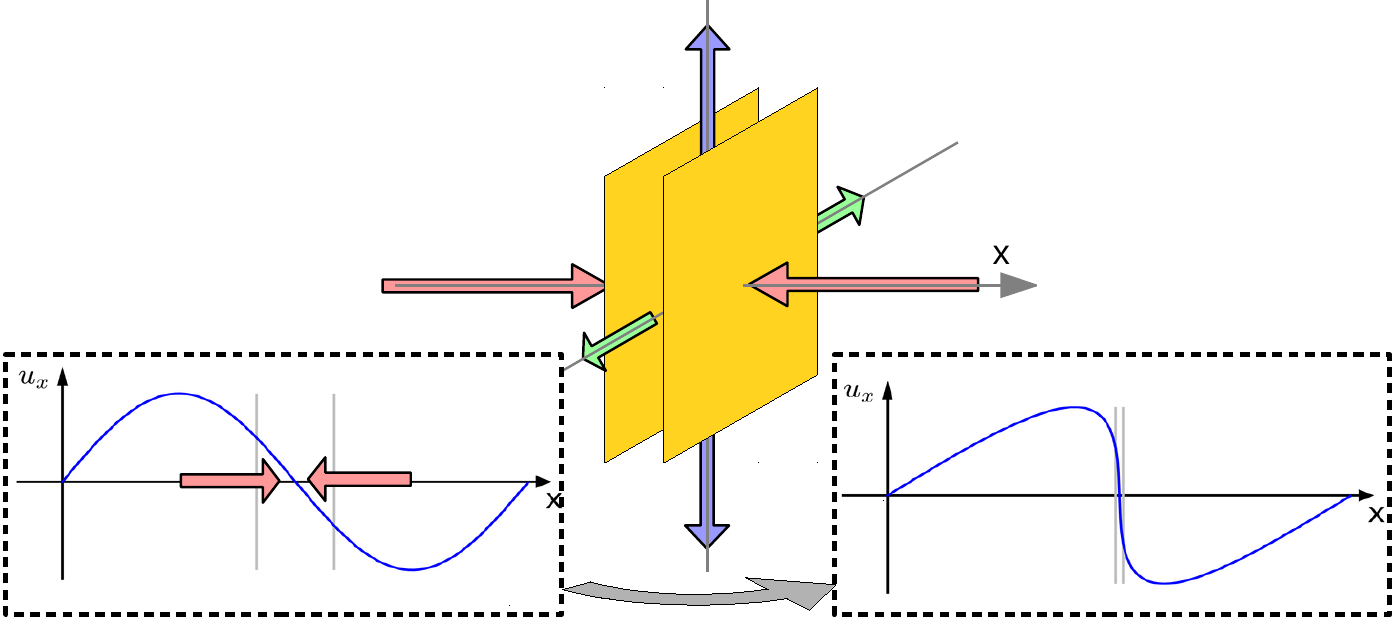}
	\caption{A simplified schematic of strain self-amplification.}
	\label{fig:strain-self-amplification}
\end{figure}

More recently, the self-amplification of straining motions has been proposed as an alternative cascade mechanism \citep{Tsinober2009, Paul2017, Carbone2020, Johnson2020}, casting doubt on the widespread view that the cascade is driven primarily by vortex stretching. Strain-rate self-amplification is illustrated in Figure \ref{fig:strain-self-amplification}. A region of enhanced strain-rate with one strong negative eigenvalue experiences a strengthening of that compressive strain-rate as faster moving fluid catches up with slower moving fluid in front of it. This decreases the spatial extent of strong compressive strain-rate. This mechanism is responsible for the finite-time singularity and shock formation in the inviscid Burgers equation, but has received much less attention in the context of Navier-Stokes. The restricted Euler equation, a dynamical system more relevant to 3D Navier-Stokes than the Burgers equation, displays a finite-time singularity that includes both strain self-amplification and vorticity stretching \citep{Vieillefosse1982}.

Indeed, more careful accounting of the relative contributions of vortex stretching and strain self-amplification in turbulent flows indicate that the latter is responsible for a larger share of the energy cascade rate. Most notably, \cite {Johnson2020} derived an exact, spatially-local relationship between the energy cascade rate and the dynamical mechanisms of vortex stretching and strain self-amplification. It is the goal of this paper to leverage this result to illuminate key aspects of the cascade as seen from the perspective of velocity gradient dynamics.

In this paper, the roles of vortex stretching and strain self-amplification in determining the energy cascade rate are elucidated. First, \S \ref{sec:theory} describes the framework for studying the turbulent energy cascade in terms of velocity gradients dynamics, enhancing and building upon the results of \citet{Johnson2020}. The connection between the restricted Euler singularity and the energy cascade is highlighted, along with the role of pressure in establishing global equalities. Simulation results highlight the relative importance of each mechanism to the energy cascade. Also included is a brief discussion of the inverse energy cascade in two-dimensional flows, demonstrating a comprehensive framework for turbulent cascades. Following that, \S \ref{sec:efficiency} builds on the concept of cascade efficiency from \cite{Ballouz2018}, examining the efficiency of the dynamical mechanisms comprising the cascade rate. This analysis reveals the relevance of restricted Euler dynamics to the energy cascade, while judging the relative suitability of an eddy viscosity hypothesis for representing the various dynamical processes involved. Then, \S \ref{sec:locality} returns to the question of scale-locality in terms of both vorticity stretching and strain self-amplification. It is shown that the concept of cascade efficiency can be examined on a scale-by-scale basis to test the de-correlation idea of \cite{Eyink2005}. Conclusions are drawn in \S \ref{sec:conclusion}.

\section{Energy cascade in terms of filtered velocity gradients}\label{sec:theory}

\subsection{Simulation database: homogeneous isotropic turbulence}
In this paper, the kinetic energy cascade is quantified via direct numerical simulation (DNS) of forced homogeneous isotropic turbulence (HIT) in a triply periodic domain. The incompressible Navier-Stokes equations,
\begin{equation}
	\frac{\partial u_i}{\partial t} + u_j \frac{\partial u_i}{\partial x_j} = - \frac{\partial p}{\partial x_i} + \nu \nabla^2 u_i + f_i,
	\hspace{0.05\linewidth}
	\frac{\partial u_j}{\partial x_j} = 0,
	\label{eq:Navier-Stokes}
\end{equation}
are solved using a pseudo-spectral method with $1024$ collocation points in each direction. The velocity field, $\mathbf{u}$, is advanced in time with a second-order Adams-Bashforth scheme, and the pressure $p$ simply enforces the divergence-free condition. The $2\sqrt{2}/3$ rule for wave number truncation is used with phase-shift dealiasing \citep{Patterson1971}. The forcing, $\mathbf{f}$, is specifically designed to maintain constant kinetic energy in the first two wave number shells. 

The simulation was initialized using a Gaussian random velocity field satisfying a model turbulent energy spectrum. The simulation was first run through a startup period to reach statistical stationarity. Then, statistics are computed over $6$ large-eddy turnover times. The Taylor-scale Reynolds number is approximately $Re_\lambda = 400$ with grid resolution $k_{\max} \eta = 1.4$. The integral length scale is about $20\%$ of the periodic box size of $2\pi$ and $L/\eta= 460$. The skewness of the longitudinal velocity gradient is $-0.58$, and the flatness of the longitudinal and transverse velocity gradients are $8.0$ and $12.4$, respectively, in reasonable agreement with previous simulations \citep{Ishihara2007}.

HIT is a very useful canonical flow to efficiently explore the energetics of small- and intermediate-scale turbulence dynamics. The analysis performed in this paper is not limited to HIT in principle, and the inertial range results are expected to be representative of a wide range of turbulent shear flows at sufficiently high Reynolds numbers.

\subsection{Velocity gradient tensor}

Vorticity stretching and strain self-amplification are dynamical processes defined via the velocity gradient tensor,
\begin{equation}
	A_{ij} = \frac{\partial u_i}{\partial x_j} = S_{ij} + \Omega_{ij},
	\hspace{0.05\linewidth}
	S_{ij} = \frac{1}{2}\left( A_{ij} + A_{ji} \right),
	\hspace{0.05\linewidth}
	\Omega_{ij} = \frac{1}{2}\left( A_{ij} - A_{ji} \right).
\end{equation}
The velocity gradient in three dimensions is a $3 \times 3$ tensor describing the variation of the three velocity components in each of the three coordinate directions. It is comprised of the strain-rate tensor, $\mathbf{S}$, which describes the rate at which fluid particles are undergoing deformation, and the rotation-rate tensor, $\boldsymbol{\Omega}$, which describes the rate at which fluid particles are undergoing solid body rotation. Mathematically speaking, the strain-rate and rotation-rate tensors are the symmetric and anti-symmetric parts of the velocity gradient tensor, respectively.

The decomposition of $\mathbf{A}$ into $\mathbf{S}$ and $\boldsymbol{\Omega}$ is foundational to the energetics of turbulent flows because viscosity resists deformation but not rotation. Thus, the rate at which kinetic energy is dissipated into thermal energy depends only on the (Frobenius norm of the) strain-rate tensor,
\begin{equation}
	\epsilon = 2 \nu S_{ij} S_{ij} = 2 \nu \| \mathbf{S} \|^2.
\end{equation}
On the other hand, the rotation-rate tensor, which may be written in terms of the vorticity vector $\boldsymbol{\omega}$,
\begin{equation}
	\Omega_{ij} = -\frac{1}{2} \epsilon_{ijk} \omega_k,
	\hspace{0.05\linewidth}
	\omega_i = - \epsilon_{ijk} \Omega_{jk},
\end{equation}
plays no direct role in the viscous dissipation of kinetic energy. However, vorticity is still essential to the dynamics of energy dissipation, as may be seen from the first relation of \citet{Betchov1956} for incompressible homogeneous turbulence,
\begin{equation}
	\langle \| \mathbf{S}\|^2 \rangle = \frac{1}{2}\langle |\boldsymbol{\omega}|^2 \rangle.
	\label{eq:Betchov-2}
\end{equation}
Angle brackets denote ensemble averaging. The Betchov relation shows that the average (or global) amount of vorticity and strain-rate is held in balance for incompressible flows, presumably by the action of the pressure as it enforces $\nabla \cdot \mathbf{u} = 0$. Thus, the significantly enhanced dissipation rates in turbulence cannot occur without equally enhanced enstrophy. At sufficiently high Reynolds numbers, this relationship holds to good approximation even for inhomogeneous flows.

For this reason, it is useful to consider (one half of) the norm of the full velocity gradient tensor,
\begin{equation}
	\frac{1}{2 }\|\mathbf{A}\|^2 = \frac{1}{4} |\boldsymbol{\omega}|^2 + \frac{1}{2} \| \mathbf{S} \|^2
\end{equation}
when considering the mean rate of dissipation. Statistically, the velocity gradient magnitude is equi-partitioned between enstrophy and dissipation. Thus, the second invariant of the velocity gradient tensor,
\begin{equation}
	Q = -\frac{1}{2} A_{ij} A_{ji} = \frac{1}{4} |\boldsymbol{\omega}|^2 - \frac{1}{2} \| \mathbf{S} \|^2,
	\label{eq:second-invariant}
\end{equation}
has an average of zero.

\begin{figure}
	\centering
	\includegraphics[page=1, width=1.0\linewidth]{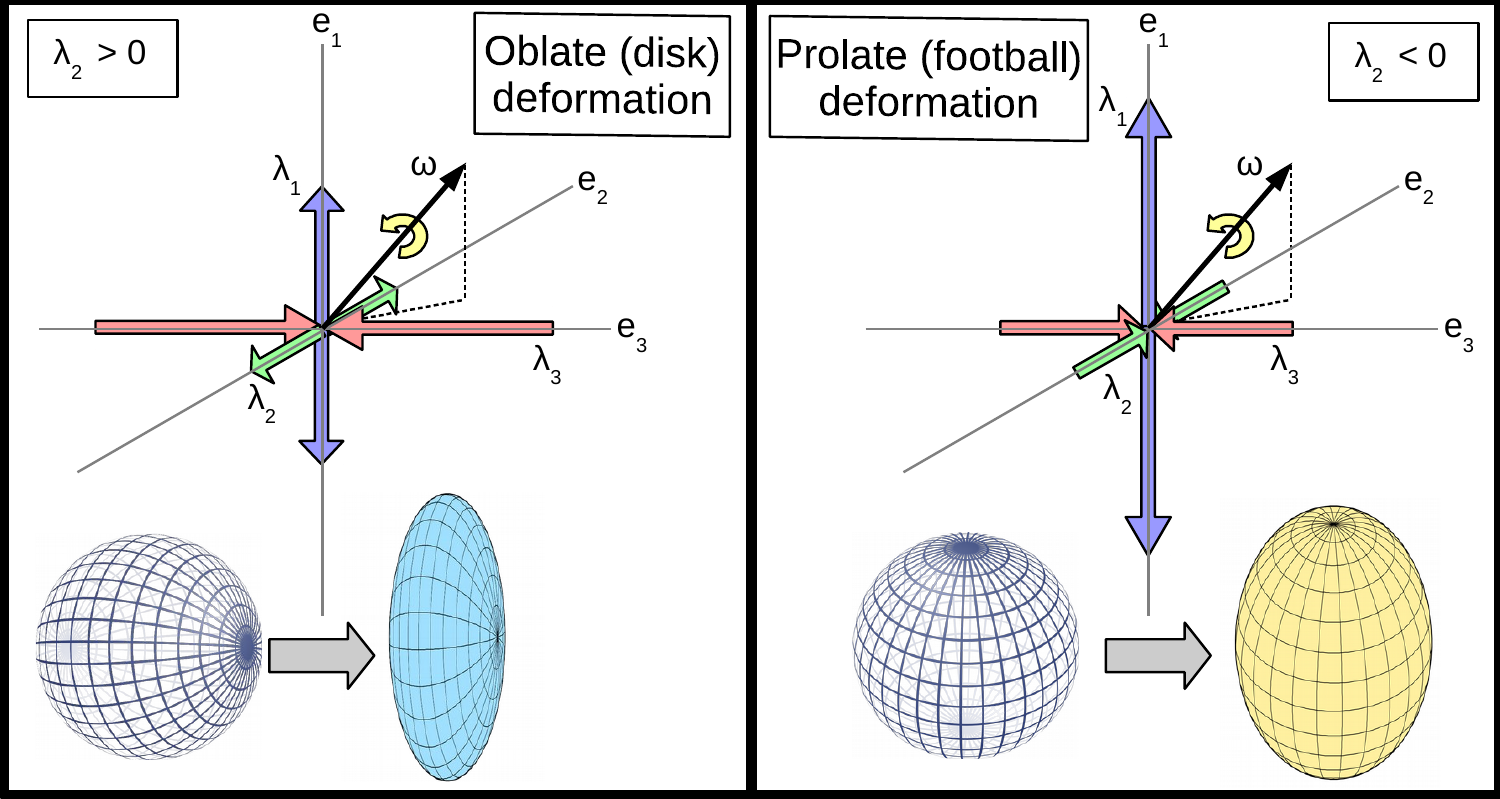}
	\caption{Depiction of the velocity gradient tensor in the strain-rate eigenframe. Velocity gradients   with two extensional eigenvalues (left) tend to deform initially spherical fluid particles into disk-like oblate spheroids. Velocity gradients with two compressive eigenvalues (right) deform spherical fluid particles to prolate spheriods, like an (American) football shape.}
	\label{fig:velocity-gradient}
\end{figure}

The strain-rate tensor, being symmetric, has 3 real eigenvalues, $\lambda_1 > \lambda_2 > \lambda_3$, associated with a set of orthogonal eigenvectors. The eigenvalues sum to zero for incompressible flows, $\lambda_1 + \lambda_2 + \lambda_3 = 0$. Therefore, the largest eigenvalue is always extensional along its eigenvector, $\lambda_1 \geq 0$, and the smallest one is always compressive, $\lambda_3 \leq 0$. The intermediate eigenvalue, $\lambda_2$, may be positive or negative. The sign of $\lambda_2$ indicates the topology of fluid particle deformations, Figure \ref{fig:velocity-gradient}, and has more important dynamical effects as discussed below.

\subsection{Velocity gradient dynamics}
Consider a Lagrangian view of turbulence as a collection of fluid particles characterized by their deformational and rotational behavior, i.e., their velocity gradient tensor. The Lagrangian evolution equation for the velocity gradient tensor is derived as the gradient of Eq. \eqref{eq:Navier-Stokes},
\begin{equation}
	\frac{DA_{ij}}{Dt} \equiv \frac{\partial A_{ij}}{\partial t} + u_k \frac{\partial A_{ij}}{\partial x_k} = - A_{ik} A_{kj} - \frac{\partial^2 p}{\partial x_i \partial x_j} + \nu \nabla^2 A_{ij} + \frac{\partial f_i}{\partial x_j}.
	\label{eq:LaVelGrad}
\end{equation}
For HIT simulations with large-scale forcing, the gradient of the force is typically negligible compared to other terms in Eq. \eqref{eq:LaVelGrad} at high Reynolds numbers. The first term on the right side of Eq. \eqref{eq:LaVelGrad}, $A_{ik} A_{kj}$, contains the dynamical mechanisms of interest: vorticity stretching/tilting/compressing and strain self-amplification/attenuation.  The second term on the right side of Eq. \eqref{eq:LaVelGrad} is the pressure Hessian tensor. It may be split into local and nonlocal parts, i.e., the isotropic and deviatoric parts, respectively \citep{Ohkitani1995},
\begin{equation}
	\frac{\partial^2 p}{\partial x_i \partial x_j}(\mathbf{x})
	=
	\underbrace{
		\frac{2}{3} Q(\mathbf{x}) \delta_{ij}
	}_{\text{isotropic, local}}
	+
	\underbrace{
		\iiint_{P.V.} d\mathbf{r} ~\frac{Q(\mathbf{x}+\mathbf{r})}{2\pi |\mathbf{r}|^3} \left( \delta_{ij} - 3 \frac{r_i r_j}{|\mathbf{r}|^2} \right)
	}_{\text{deviatoric, nonlocal}}.
	\label{eq:pressure-Hessian}
\end{equation}
The isotropic part of this tensor is given by the pressure Poisson equation (from the divergence of Eq. \eqref{eq:Navier-Stokes}),
\begin{equation}
	\nabla^2 p = 2 Q,
	\label{eq:pressure-Poisson}
\end{equation}
which is the equation locally enforcing the divergence free condition along the Lagrangian trajectory. The deviatoric part of the pressure Hessian, given by the nonlocal integral in Eq. \eqref{eq:pressure-Hessian}, represents the action of the pressure to enforce $\nabla \cdot \mathbf{u} = 0$ at nearby points in the flow. Splitting the pressure Hessian in this way enables a decomposition of the Lagrangian velocity gradient evolution, Eq. \eqref{eq:LaVelGrad}, into the autonomous dynamics of a fluid particle and the influence of other nearby fluid particles,
\begin{equation}
	\frac{DA_{ij}}{Dt}
	=
	- \underbrace{ \left(
		A_{ik}A_{kj} - \frac{1}{3} A_{mn} A_{nm} \delta_{ij}
		\right)
	}_{\text{autonomous dynamics}}
	-
	\underbrace{ \left( 
		\frac{\partial^2 p}{\partial x_i x_j} - \frac{1}{3} \nabla^2 p \delta_{ij}
		\right)
		+ \nu \nabla^2 A_{ij}
	}_{\text{nearby particle interactions}}.
	\label{eq:LaVelGrad-split}
\end{equation}

Significant insight may be gained into velocity gradient dynamics by focusing on its autonomous dynamics using the restricted Euler equation,
\begin{equation}
	\frac{DA_{ij}}{Dt}
	=
	- A_{ik}A_{kj} - \frac{1}{3} A_{mn} A_{nm} \delta_{ij}.
	\label{eq:restricted-Euler}
\end{equation}
By neglecting the interaction with the velocity gradients of other surrounding fluid particles, the restricted Euler equation represents a dramatic mathematical simplification of turbulent flows while retaining some key dynamical processes. In fact, rigorous mathematical results are possible \citep{Vieillefosse1982, Vieillefosse1984, Cantwell1992}.

The most salient feature of solutions to the restricted Euler equation is a finite-time singularity for (almost) all initial conditions. The cause of this singularity may be readily seen by considering the equation for the velocity gradient norm in restricted Euler dynamics,
\begin{equation}
	\frac{D}{Dt}\left( \frac{1}{2} \| \mathbf{A} \|^2 \right) = \frac{1}{4} \omega_i S_{ij} \omega_j - S_{ij} S_{jk} S_{ki}.
	\label{eq:restricted-Euler-norm}
\end{equation}
The two terms on the right side of Eq. \eqref{eq:restricted-Euler-norm} represent production of velocity gradient magnitude by vorticity stretching and strain self-amplification, respectively.

The local rate of velocity gradient production by vorticity stretching/compression,
\begin{equation}
	P_\omega = \frac{1}{4} \omega_i S_{ij} \omega_j = \frac{1}{4} |\boldsymbol{\omega}|^2 \sum_{i=1}^{3} \lambda_i \cos^2(\theta_{\omega,i}),
	\label{eq:vorticity-stretching}
\end{equation}
depends strongly on how the vorticity vector aligns in the strain-rate eigenframe, Figure \ref{fig:velocity-gradient}. Here, $\theta_{\omega,i}$ is the angle between the vorticity and the i$^{th}$ eigenvector of the strain-rate tensor. The local enstrophy, and hence the velocity gradient magnitude, is increased when the vorticity vector aligns more with eigenvectors having positive (extensional) eigenvalues. Velocity gradient production rate via strain self-amplification is,
\begin{equation}
	P_s = - S_{ij} S_{jk} S_{ki} = - 3 \lambda_1 \lambda_2 \lambda_3.
	\label{eq:strain-self-amplification}
\end{equation}
The strain-rate amplifies itself when $\lambda_2 > 0$ but self-attenuates when $\lambda_2 < 0$. It is no surprise, then, that restricted Euler solutions tend toward a state with two positive strain-rate eigenvalues, $\lambda_1 = \lambda_2 = -\frac{1}{2}\lambda_3$, as they approach the finite-time singularity. Furthermore, the vorticity aligns with the eigenvector associated with the intermediate eigenvalue, $\lambda_2$, in this approach to singularity. Thus, both strain self-amplification and vorticity stretching together drive the finite-time singularity in restricted Euler.

However, strain self-amplification dominates vorticity stretching in the restricted Euler singularity, as can be demonstrated by considering the dynamics in terms of the second and third invariants. The second invariant, $Q$, is defined in Eq. \eqref{eq:second-invariant} as the difference between the square magnitudes of vorticity and strain-rate. The third invariant,
\begin{equation}
	R = -\frac{1}{3} A_{ij} A_{jk} A_{ki}
	= -\frac{1}{3} S_{ij} S_{jk} S_{ki} - \frac{1}{4} \omega_i S_{ij} \omega_j
	= \frac{1}{3} P_s - P_\omega
	\label{eq:third-invariant}
\end{equation}
is the difference between (one third of) the strain self-amplification rate and the vorticity stretching rate. The restricted Euler equation may be written in terms of $Q$ and $R$ as
\begin{equation}
	\frac{DQ}{Dt} = - 3R,
	\hspace{0.05\linewidth}
	\frac{DR}{Dt} = \frac{2}{3}Q^2.
\end{equation}
Thus, $DR/Dt \geq 0$ always. Restricted Euler solutions evolve toward positive $R$ which is characterized by more strain self-amplification than vorticity stretching. \citet{Vela-Martin2020} illuminates a subset of this behavior by looking at the dynamics of $\lambda_2$. Further, as $R$ becomes positive, $DQ/Dt < 0$ and the second invariant decreases. The result is that the finite-time singularities in restricted Euler solutions are characterized by $R \gg 0$ and $Q \ll 0$. That is, the strain-rate blows up faster than the vorticity.

Incompressible Navier-Stokes dynamics, with the nonlocal pressure Hessian and viscous term reintroduced, do not exhibit the finite-time singularity seen for the restricted Euler equation. Instead, there is a balance between vorticity and strain-rate as already discussed with Eq. \eqref{eq:Betchov-2}. \cite{Betchov1956} provided another relation for incompressible homogeneous turbulence, namely,
\begin{equation}
-\langle S_{ij} S_{jk} S_{ki} \rangle = \frac{3}{4}\langle \omega_i S_{ij} \omega_j \rangle,
\hspace{0.05\linewidth}
\text{or}
\hspace{0.05\linewidth}
\langle P_s \rangle = 3 \langle P_\omega \rangle,
\label{eq:Betchov-3}
\end{equation}
which means that the average value of $R$ is also zero, as enforced by the pressure via the constraint $\nabla \cdot \mathbf{u} = 0$. As with Eq. \eqref{eq:Betchov-2}, Eq. \eqref{eq:Betchov-3} is approximately true for high Reynolds number inhomogeneous flows. What this means is that, while the pressure enforces an equal partition between dissipation and enstrophy, it also enforces that the strain-rate self-amplification produces stronger velocity gradients at three times the rate of vorticity stretching.

The significance of the restricted Euler equation is that its signature features are unmistakably observed in the statistics of experiments and direct numerical simulations of the Navier-Stokes equations. Indeed, turbulent flows show a tendency of the vorticity to align with the strain-rate eigenvector associated with the intermediate eigenvalue, $\lambda_2$, which tends to be positive much more often than it is negative \citep{Kerr1987, Ashurst1987, Tsinober1992, Lund1994, Mullin2006, Gulitski2007}. Note that the alignment of vorticity is similar to, but slightly different from, the alignment of material lines in a turbulent flow \citep{Luthi2005, Holzner2010, Johnson2016a, Johnson2017a}, a subtle point to remember in the context of \citet{Taylor1938}. An important difference between vorticity and material lines is that the vorticity is an active feedback on the strain-rate and evolves on the same time scale, so that while the vorticity typically tilts toward the strain-rates most stretching eigenvalue, it is more successful in aligning with direction in which that eigenvector previously pointed \citep{Xu2011}. In addition, turbulence is characterized by enhanced probabilities of strong velocity gradients along the Vieillefosse manifold, $4 Q^3 + 27 R^2 = 0$ in the fourth quadrant, which is an attracting manifold along which the finite-time singularity occurs for restricted Euler \citep{Cantwell1993, Soria1994, Chong1998, Nomura1998, Ooi1999, Gulitski2007, Elsinga2010}. While the nonlocal part of the pressure Hessian tensor, along with viscous effects, are important for describing turbulence quantitatively, many of the unique qualitative features of turbulent velocity gradients can be connected to restricted Euler dynamics.

In summary, velocity gradients in a fluid undergoing nonlinear self-advection ($\mathbf{u}\cdot\nabla\mathbf{u}$) naturally strengthen via vorticity stretching and strain self-amplification. Of these two dynamical processes, strain self-amplification is three times as strong on average, as enforced by the non-local action of the pressure. For more discussion about velocity gradient dynamics, as well as measurement and modelling, the reader is referred to \citep{Wallace2009, Tsinober2009, Meneveau2011}. Next, the energy cascade is introduced using a spatial filtering formulation. As will be shown, this framework allows for directly relating velocity gradient dynamics to the energy cascade.

\subsection{Spatial filtering and the energy cascade}

\begin{figure}
	\centering
	\includegraphics[width=0.49\linewidth]{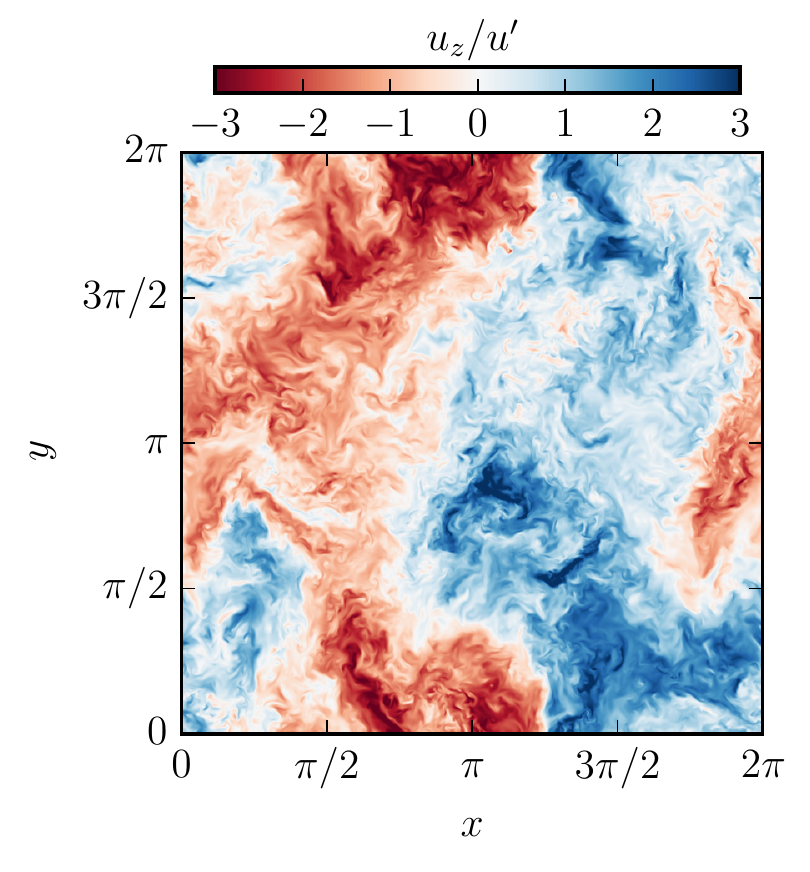}
	\includegraphics[width=0.49\linewidth]{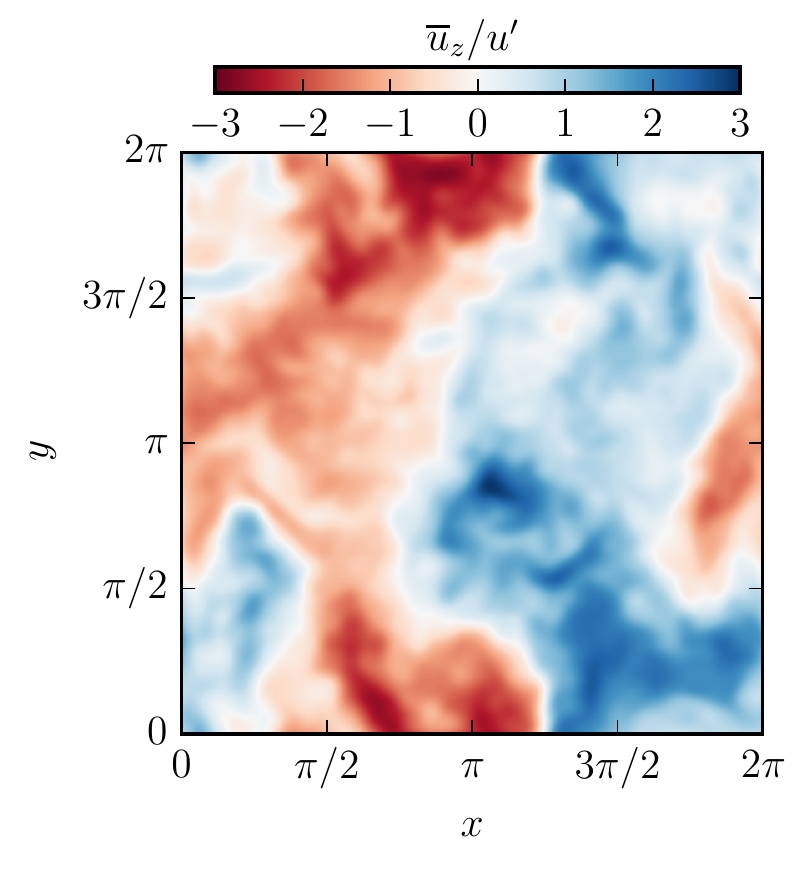}
	\caption{The fluid velocity in the $z$ direction on an $xy$ plane in the HIT simulation: (a) unfiltered, (b) filtered at $\ell = 25\eta$ using a Gaussian filter, Eq. \eqref{eq:Gaussian-kernel}.}
	\label{fig:filter-velocity}
\end{figure}

A spatial low-pass filter,
\begin{equation}
	\overline{a}^\ell(\mathbf{x}) = \iiint d\mathbf{r}~ G_\ell(\mathbf{r}) a(\mathbf{x} + \mathbf{r}),
	\hspace{0.05\linewidth}
	\widehat{\overline{a}^\ell}(\mathbf{k}) = \widehat{G_\ell}(\mathbf{k}) \widehat{a}(\mathbf{k})
\end{equation}
retains features of a field $a(\mathbf{x})$ that are larger than $\ell$ while removing features smaller than $\ell$ \citep{Leonard1975, Germano1992}. An example velocity field from the DNS of HIT is shown in Figure \ref{fig:filter-velocity}, before and after the application of a spatial filter. The spatial filter may be interpreted as a form of local averaging, weighted by the filter kernal $G_\ell(\mathbf{r})$. Following \citet{Germano1992}, the generalized second moment is the difference of the filtered product and the product of two filtered fields,
\begin{equation}
	\tau_\ell(a,b) = \overline{a b}^\ell - \overline{a}^\ell \overline{b}^\ell.
	\label{eq:generalized-second-moment}
\end{equation}
It is a generalized covariance of small-scale activity (smaller than $\ell$).

The kinetic energy can be thus defined as the sum of large-scale and small-scale energies,
\begin{equation}
	\tfrac{1}{2}\overline{u_i u_i}^\ell = \tfrac{1}{2}\overline{u}_i^\ell \overline{u}_i^\ell + \tfrac{1}{2}\tau_\ell(u_i, u_i).
\end{equation}
If the filter kernel is positive semi-definite, i.e., $G_\ell(\mathbf{r}) \geq 0$ for all $\mathbf{r}$, then the subfilter-scale energy is positive everywhere, $\tau(u_i, u_i)(\mathbf{x}) \geq 0$ for all $\mathbf{x}$ \citep{Vreman1994}.

Applying a low-pass filter to the incompressible Navier-Stokes equations, Eq. \eqref{eq:Navier-Stokes}, results in an evolution equation for the filtered velocity field,
\begin{equation}
	\frac{\partial \overline{u}_i^\ell}{\partial t} + \overline{u}_j ^\ell\frac{\partial \overline{u}_i^\ell}{\partial x_j} = - \frac{\partial \overline{p}^\ell}{\partial x_i} + \nu \nabla^2 \overline{u}_i^\ell - \frac{\partial \tau_\ell(u_i, u_j)}{\partial x_j} + \overline{f}_i^\ell.
	\label{eq:filtered-Navier-Stokes}
\end{equation}
The main difference with the unfiltered Navier-Stokes equation, Eq. \eqref{eq:Navier-Stokes}, is the introduction of the subfilter stress tensor's divergence on the right side. The dynamical equation for the large-scale kinetic energy directly follows from multiplication of Eq. \eqref{eq:filtered-Navier-Stokes} by $\overline{u}_i^\ell$,
\begin{equation}
	\frac{\partial \left(\tfrac{1}{2}\overline{u}_i^\ell \overline{u}_i^\ell\right)}{\partial t}
	+  \frac{\partial \Phi_j^\ell}{\partial x_j}
	= 
	\overline{u}_i^\ell \overline{f}_i^\ell
	+ \tau_\ell(u_i,u_j) \overline{S}_{ij}^\ell
	- 2 \nu \overline{S}_{ij}^\ell \overline{S}_{ij}^\ell,
	\label{eq:resolved-energy}
\end{equation}
\begin{equation*}
	\text{where}
	\hspace{0.05\linewidth}
	\Phi_{j}^\ell = \tfrac{1}{2}\overline{u}_i^\ell \overline{u}_i^\ell \overline{u}_j^\ell
	+ \overline{p}^\ell \overline{u}_j^\ell
	- 2 \nu \overline{u}_i^\ell \overline{S}_{ij}^\ell
	+ \overline{u}_i^\ell \tau_\ell(u_i,u_j).
\end{equation*}
The spatial transport of large-scale kinetic energy, $\boldsymbol{\Phi}^\ell$, occurs due to (i) advection by large-scale motions, $\tfrac{1}{2}\overline{u}_i^\ell \overline{u}_i^\ell \overline{u}_j^\ell$, (ii) large-scale pressure transport, $\overline{p}^\ell \overline{u}_j^\ell$, (iii) viscous diffusion, $- 2 \nu \overline{u}_i^\ell \overline{S}_{ij}^\ell$, and (iv) mixing by subfilter-scale motions $\overline{u}_i^\ell \tau_\ell(u_i,u_j)$.

More importantly for the present topic, there are three source/sink terms. First, large-scale kinetic energy is produced though work done by the applied forcing function, $\overline{u}_i^\ell \overline{f}_i^\ell$. Second, the subfilter stress may add or remove energy depending on its alignment with the filtered strain-rate tensor, $\tau_\ell(u_i,u_j) \overline{S}_{ij}^\ell$. Finally, the direct dissipation of large-scale kinetic energy by viscosity, $- 2 \nu \overline{S}_{ij}^\ell \overline{S}_{ij}^\ell$, is typically negligible if $\ell \gg \eta$.

An equation for total kinetic energy, $\tfrac{1}{2} \overline{u_i u_i}^\ell$, is constructed by multiplying Eq. \eqref{eq:Navier-Stokes} by $u_i$ and then filtering. Equation \eqref{eq:resolved-energy} is subtracted from the resulting equation to form the transport equation for the small-scale energy,
\begin{equation}
	\frac{\partial \left( \tfrac{1}{2}\tau_\ell(u_i,u_j) \right)}{\partial t} + \frac{\partial \phi_j^\ell}{\partial x_j}
	=
	\tau_\ell\left(u_i, f_i\right)
	- \tau_\ell\left(u_i, u_j\right) \overline{S}_{ij}^\ell
	- 2 \nu ~\tau_\ell\left(S_{ij}, S_{ij}\right),
	\label{eq:residual-energy}
\end{equation}
\begin{equation*}
	\text{where}
	\hspace{0.05\linewidth}
	\phi_j^\ell
	= \tfrac{1}{2} \tau_\ell\left(u_i, u_i\right) \overline{u}_j^\ell + \tfrac{1}{2}\tau_\ell\left(u_i, u_i, u_j\right) + \tau_\ell\left(p, u_j\right) - 2 \nu ~\tau_\ell\left(u_i, S_{ij}\right).
\end{equation*}
Spatial transport of small-scale kinetic energy, $\boldsymbol{\phi}^\ell$, is done by (i) large-scale advection, $\tfrac{1}{2} \tau_\ell\left(u_i, u_i\right) \overline{u}_j$, (ii) small-scale mixing, $\tfrac{1}{2}\tau_\ell\left(u_i, u_i, u_j\right)$, (iii) small-scale pressure transport, $\tau_\ell\left(p, u_j\right)$, and (iv) viscous diffusion, $2 \nu ~\tau_\ell\left(u_i, S_{ij}\right)$. The second transport term is a generalized third moment \citep{Germano1992},
\begin{equation}
	\tau_\ell(a, b, c) = \overline{a b c}^\ell - \overline{a}^\ell \tau_\ell(b,c) - \overline{b}^\ell \tau_\ell(a,c) - \overline{c}^\ell \tau_\ell(a,b) - \overline{a}^\ell \overline{b}^\ell \overline{c}^\ell.
\end{equation}

There are three sources/sinks in Eq. \eqref{eq:residual-energy}, which have the following significance. The first source is direct forcing of the small scales, $\tau_\ell\left(u_i, f_i\right)$, which is typically negligible for $\ell \ll L$. The second is the same term, $- \tau_\ell\left(u_i, u_j\right) \overline{S}_{ij}^\ell$, as appears with opposite sign in the large-scale energy equation, Eq. \eqref{eq:resolved-energy}. This term represents the rate at which energy is transferred from motions of size larger than $\ell$ to motions of size smaller than $\ell$. Thus, the energy cascade rate is defined as \citep{Leonard1975, Germano1992, Meneveau2000},
\begin{equation}
	\Pi^\ell = - \mathring{\tau}_\ell\left(u_i, u_j\right) \overline{S}_{ij}^\ell,
\end{equation}
where the overset circle indicates the deviatoric component of the tensor,
\begin{equation}
	\mathring{\tau}_\ell(u_i, u_j) = \tau_\ell(u_i, u_j) - \tfrac{1}{3} \tau_\ell(u_k, u_k) \delta_{ij}.
\end{equation}
The isotropic part of the subfilter stress tensor does not contribute to the energy cascade rate, $\Pi^\ell$, because the trace of $\overline{\mathbf{S}}^\ell$ is zero due to incompressibility. Positive cascade rate indicates energy transfer from large to small scales and negative rate indicates backscatter or inverse cascade. It may be interpreted as the rate at which large-scale motions do work on small-scale motions \citep{Ballouz2018}.

The sign of $\Pi^\ell$ depends on how the eigenvectors of the subfilter stress tensor aligns with those of the filtered strain-rate tensor \citep{Ballouz2020}. This may be made explicit by writing the cascade rate in terms of eigenvalues of the filtered strain-rate, $\lambda_i$, and deviatoric subfilter stress tensor, $\mu_j$,
\begin{equation}
\Pi^\ell
= \sum_{i=1}^{3} \sum_{j=1}^{3} \lambda_i^\ell \mu_j^\ell \cos^2(\theta_{ij}^\ell).
\label{eq:Pi-definition}
\end{equation}
Here, $\theta_{ij}$ is the angle between the i$^{th}$ eigenvector of the filtered strain-rate and j$^{th}$ eigenvector of the subfilter stress tensor.

\begin{figure}
	\includegraphics[width=0.47\linewidth]{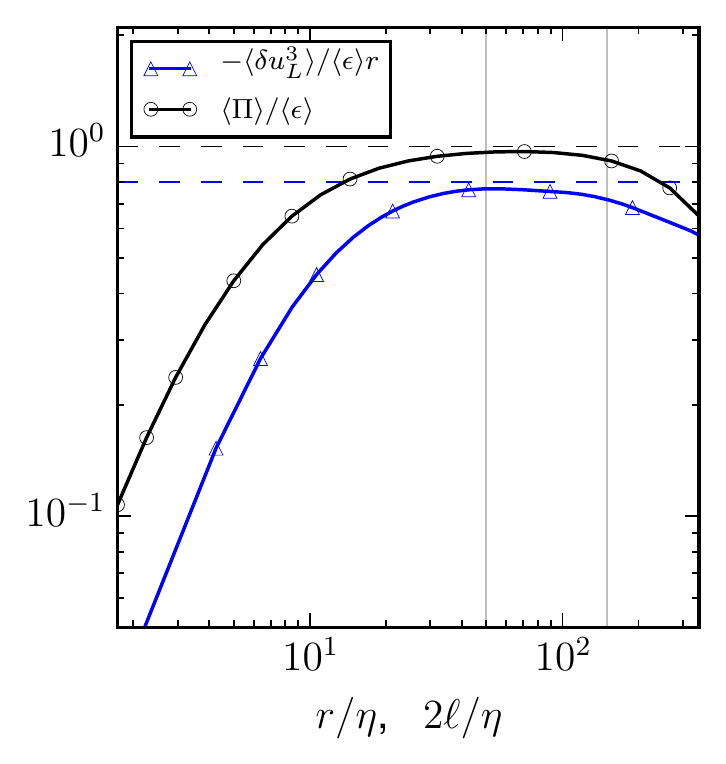}
	\includegraphics[width=0.517\linewidth]{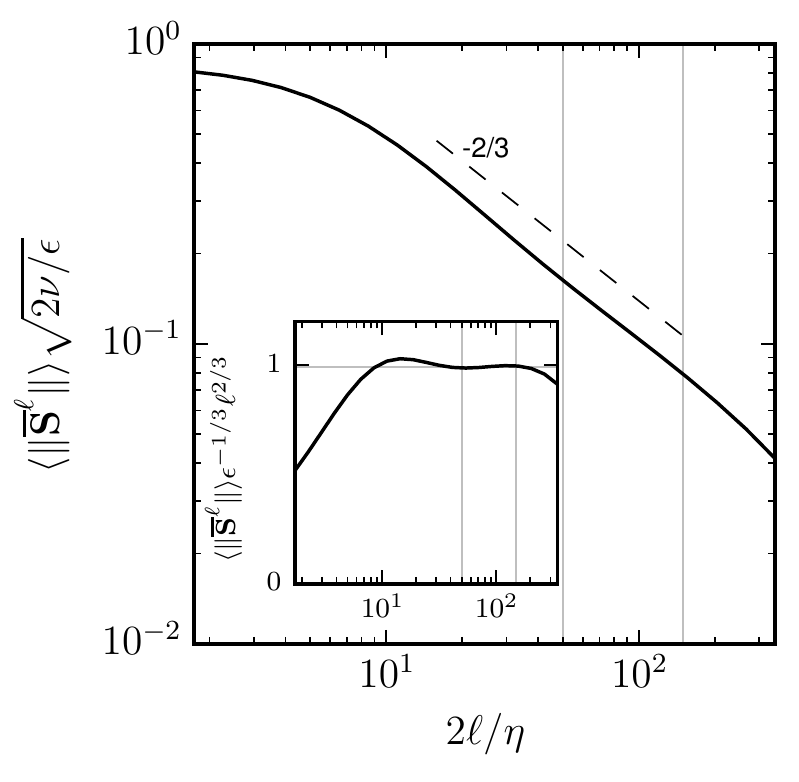}
	\caption{(a) The pre-multiplied third-order longitudinal structure function alongside the mean interscale energy transfer. (b) The average norm of the filtered strain-rate tensor as function of filter width, $\ell$. A $-2/3$ power law is consistent with inertial range behavior. The inset is premultiplied by $\ell^{2/3}$.
		In both sub-panels, the two vertical gray lines are at $2\ell/\eta = 50$ and $150$, indicating the approximate inertial range of scales. The integral length scale is $2 L / \eta = 920$.
	}
	\label{fig:inertial_range}
\end{figure}

Averaging Eqs. \eqref{eq:resolved-energy} and \eqref{eq:residual-energy} for a stationary, homogeneous flow yields, respectively,
\begin{equation}
	0
	= 
	\left\langle \overline{u}_i^\ell \overline{f}_i^\ell \right\rangle
	+ \left \langle \tau_\ell(u_i,u_j) \overline{S}_{ij}^\ell \right\rangle
	- 2 \nu \left\langle \overline{S}_{ij}^\ell \overline{S}_{ij}^\ell \right\rangle,
	\label{eq:resolved-energy-avg}
\end{equation}
\begin{equation}
	0
	=
	\left\langle \tau_\ell\left(u_i, f_i\right) \right\rangle
	- \left\langle \tau_\ell\left(u_i, u_j\right) \overline{S}_{ij}^\ell \right\rangle
	- 2 \nu \left\langle \tau_\ell\left(S_{ij}, S_{ij}\right)\right\rangle.
	\label{eq:residual-energy-avg}
\end{equation}
In the inertial range of turbulence, $\eta \ll \ell \ll L$, the viscous dissipation of the large scales and forcing of the small scales may be neglected,
\begin{equation}
	\left\langle u_i f_i \right\rangle
	\approx 
	\left\langle \overline{u}_i^\ell \overline{f}_i^\ell \right\rangle
	\approx
	- \left \langle \tau_\ell(u_i,u_j) \overline{S}_{ij}^\ell \right\rangle
	\approx
	2 \nu \left\langle \tau_\ell\left(S_{ij}, S_{ij}\right)\right\rangle
	\approx
	2 \nu \left\langle S_{ij} S_{ij}  \right\rangle.
	\label{eq:inertial-range}
\end{equation}
More simply, the inertial range is the range length scales, $\ell$, for which $\left\langle \Pi \right\rangle \approx \left\langle \epsilon \right\rangle$. 
Figure \ref{fig:inertial_range}a tests for scales where this relation approximately holds.

\subsection{Filtered velocity gradients}

Velocity increments,
\begin{equation}
	\delta u_i(\mathbf{r}; \mathbf{x}) = u_i(\mathbf{x}+\mathbf{r}) - u_i(\mathbf{x})
\end{equation}
and structure functions are one of the prominent tools of classical turbulence theory. Filtered velocity gradients are intimately related to velocity increments \citep{Eyink1995},
\begin{equation}
	\overline{A}_{ij}^\ell(\mathbf{x}) \equiv \frac{\partial \overline{u}_i}{\partial x_j}(\mathbf{x}) 
	= \iiint d\mathbf{r}~ \left.\frac{\partial G}{\partial r_j}\right|_{\mathbf{r}} \delta u_i(\mathbf{r}, \mathbf{x})
	= \int_{0}^{\infty} d\rho~ \frac{dG}{d\rho} \left[ \oiint dS ~\frac{r_j}{|\mathbf{r}|} \delta u_i(\mathbf{r}; \mathbf{x}) \right]
	\label{eq:filtered-velocity-gradient}
\end{equation}
where the filter kernel is assumed to be spherically symmetric. That is, the filtered velocity gradient is an averaging of velocity increments weighted by the gradient of the filter kernel. For a top-hat filter kernel, the gradient of the filter kernel is a Dirac delta function, so the filtered velocity gradient components are directionally-weighted averages of velocity increments on a spherical shell at $|\mathbf{r}| = \ell$. Velocity gradients for other filter kernel shapes also average across increments $|\mathbf{r}|$ near $\ell$. Thus, the scaling of filtered velocity gradients in the inertial range can be deduced from the scaling of velocity increments,
\begin{equation}
	\|\overline{\mathbf{A}}^\ell\| \sim \frac{\delta u}{\ell} \sim \ell^{-2/3},
	\label{eq:filtered-velocity-gradient-scaling}
\end{equation}
modulo intermittency corrections. Filtered velocity gradients contain information about velocity increments in the flow,  arranged so as to illuminate the local topology of the flow at scale $\ell$. That is, filtered velocity gradients can be decomposed into rotation and deformation by considering the filtered vorticity vector and filtered strain-rate tensor,
\begin{equation}
	\overline{A}_{ij}^\ell = \overline{S}_{ij}^\ell + \overline{\Omega}_{ij}^\ell,
	\hspace{0.05\linewidth}
	\overline{S}_{ij}^\ell = \frac{1}{2} \left( \overline{A}_{ij}^\ell + \overline{A}_{ji}^\ell \right),
	\hspace{0.05\linewidth}
	\overline{\Omega}_{ij}^\ell = \frac{1}{2} \left( \overline{A}_{ij}^\ell - \overline{A}_{ji}^\ell \right),
\end{equation}
\begin{equation}
	\overline{\Omega}_{ij}^\ell = - \frac{1}{2} \epsilon_{ijk} \overline{\omega}_k^\ell,
	\hspace{0.05\linewidth}
	\overline{\omega}_i^\ell = - \epsilon_{ijk} \overline{\Omega}_{jk}^\ell,
\end{equation}
\begin{equation}
	\frac{1}{2} \| \overline{\mathbf{A}}^\ell \|^2 = \frac{1}{4} | \overline{\boldsymbol{\omega}}^\ell |^2 + \frac{1}{2}\| \overline{\mathbf{S}}^\ell \|^2,
	\hspace{0.05\linewidth}
	\left\langle \| \overline{\mathbf{S}}^\ell \|^2 \right\rangle = \frac{1}{2} \left\langle |\overline{\boldsymbol{\omega}}^\ell|^2 \right\rangle
	\label{eq:filtered-Betchov-2}
\end{equation}
The power-law scaling of the filtered strain-rate norm is shown in Figure \ref{fig:inertial_range}b as another indication of what filter widths may be considered to be in the inertial range.

\begin{figure}
	\centering
	\includegraphics[width=0.49\linewidth]{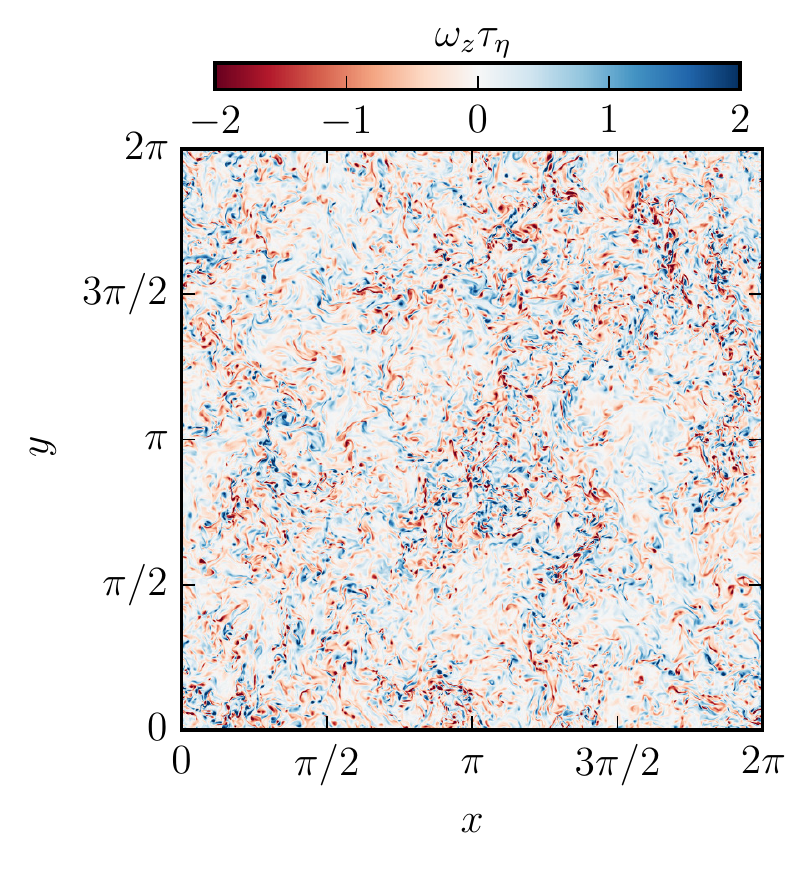}
	\includegraphics[width=0.49\linewidth]{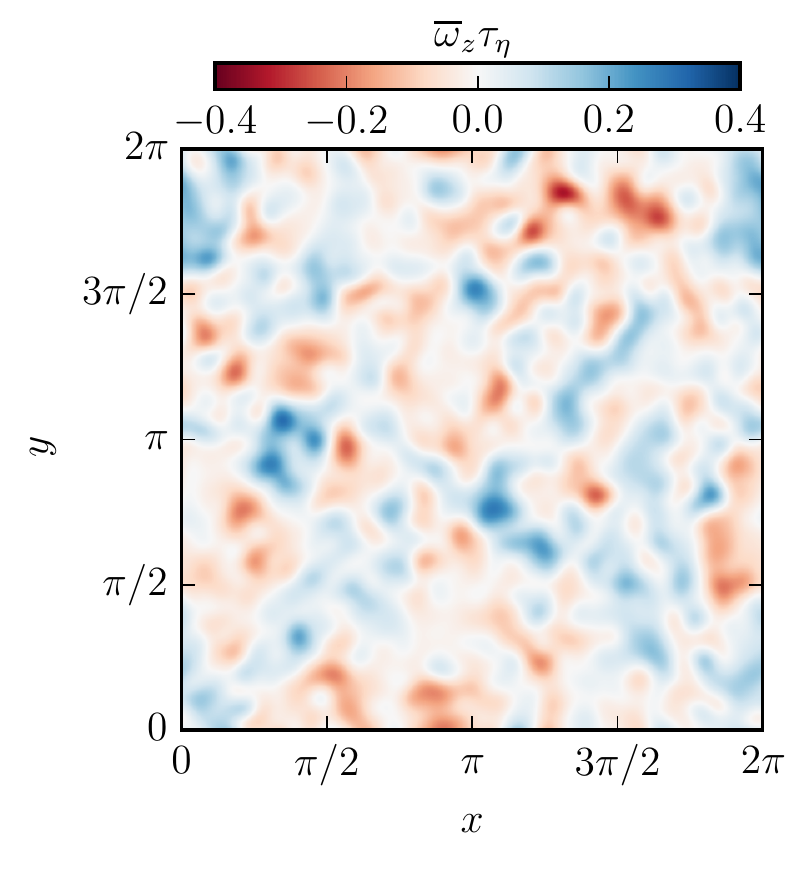}
	\caption{The $z$ component of vorticity along the same $xy$ plane from Figure \ref{fig:filter-velocity}: (a) unfiltered, (b) filtered at $\ell = 25 \eta$ using a Gaussian filter, Eq. \eqref{eq:Gaussian-kernel}.}
	\label{fig:filter-vorticity}
\end{figure}

Figure \ref{fig:filter-vorticity} shows vorticity and filtered vorticity fields from the same snapshot used for Figure \ref{fig:filter-velocity}. It is evident that the vorticity is predominantly organized at the smallest scales of motion, near $\eta$. This is true of the velocity gradient tensor in general. However, the filtered velocity gradient, as demonstrated for the filtered vorticity, is primarily organized at a scale near the filter width. Thus, the filtered velocity gradient provides a good definition of fluid motions at scale $\ell$.

Like the filtered velocity gradient, the subfilter stress tensor may also be written as a local averaging of velocity increments \citep{Constantin1994, Eyink1995},
\begin{multline}
	\tau_\ell(u_i, u_j) = 
	\left[ \iiint d\mathbf{r}~ G_\ell(\mathbf{r}) \delta u_i(\mathbf{r}; \mathbf{x}) \delta u_j(\mathbf{r}; \mathbf{x}) \right] \\
	- \left[ \iiint d\mathbf{r}~ G_\ell(\mathbf{r}) \delta u_i(\mathbf{r}; \mathbf{x})\right]
	\left[ \iiint d\mathbf{r}~ G_\ell(\mathbf{r}) \delta u_j(\mathbf{r}; \mathbf{x})\right].
\end{multline}
In this way, it can be directly shown that
\begin{equation}
	\tau_\ell(u_i, u_j) \sim \delta u^2,
	\hspace{0.03\linewidth}
	\text{and}
	\hspace{0.03\linewidth}
	\overline{S}_{ij}^\ell \sim \delta u / \ell,
	\hspace{0.03\linewidth}
	\text{therefore}
	\hspace{0.03\linewidth}
	\Pi^\ell \sim \delta u^3 / \ell
\end{equation}
so that the inertial range equation, from Eq. \eqref{eq:inertial-range}, in the form $\langle \Pi^\ell \rangle = \langle \epsilon \rangle$ is in some ways analogous to the celebrated four-fifths law of \cite{Kolmogorov1941b}, i.e., $\langle \delta u_L^3(r) \rangle = - \tfrac{4}{5} \langle \epsilon \rangle r$. The similarity between these two is highlighted in Figure \ref{fig:inertial_range}a, which includes the pre-multiplied third-order longitudinal structure function alongside $\langle\Pi\rangle / \langle \epsilon \rangle$. These two are similar diagnostics of inertial range scales. 

\subsection{Filtered velocity gradient dynamics}

Definitively linking the energy cascade rate with vorticity stretching and strain self-amplification is naturally done by considering the dynamics of filtered velocity gradients. The gradient of filtered Navier-Stokes, Eq. \eqref{eq:filtered-Navier-Stokes},
\begin{equation}
		\frac{\overline{D}~\overline{A}_{ij}^\ell}{\overline{Dt}}
		=
		- \underbrace{ \left(
			\overline{A}_{ik}^\ell \overline{A}_{kj}^\ell - \frac{1}{3} \overline{A}_{mn}^\ell \overline{A}_{nm}^\ell \delta_{ij}
			\right)
		}_{\text{autonomous dynamics}}
		-
			\left( 
			\frac{\partial^2 \overline{p}^\ell}{\partial x_i x_j} - \frac{1}{3} \nabla^2 \overline{p}^\ell \delta_{ij}
			\right)
			+ \nu \nabla^2 \overline{A}_{ij}^\ell
			- \frac{\partial \tau_\ell(u_i, u_k)}{\partial x_j \partial x_k}
		\label{eq:filtered-LaVelGrad-split}
\end{equation}
gives the dynamical evolution of filtered velocity gradients along filtered Lagrangian trajectories, $\overline{D}/\overline{Dt} = \partial/\partial t + \overline{\mathbf{u}}^\ell\cdot\nabla$. The difference from unfiltered velocity gradients, Eq. \eqref{eq:LaVelGrad-split}, is the gradient of the subfilter scale force. The dynamical consequence of the subfilter scale force is subtle, and the restricted Euler-like features observed for unfiltered velocity gradients are also seen for filtered ones \citep{Danish2018}. In other words, the autonomous dynamics of filtered velocity gradients are very influential in setting statistical trends in turbulent flows, as previously seen for unfiltered gradients.

The restricted Euler equation for filtered velocity gradient dynamics, in terms of the filtered velocity gradient norm,
\begin{equation}
	\frac{D}{Dt}\left( \frac{1}{2} \| \overline{\mathbf{A}}^\ell \|^2 \right)
	= P_{\overline{\omega}} + P_{\overline{s}}
	= \frac{1}{4} \overline{\omega}_i^\ell \overline{S}_{ij}^\ell \overline{\omega}_j^\ell - \overline{S}_{ij}^\ell \overline{S}_{jk}^\ell \overline{S}_{ki}^\ell,
	\label{eq:filtered-restricted-Euler-norm}
\end{equation}
highlights the role of vorticity stretching and strain self-amplification in increasing the magnitude of filtered velocity gradients. Recall from Eq. \eqref{eq:filtered-Betchov-2} that the filtered vorticity and strain-rate are also held in statistical equi-partition by the zero divergence condition, $\nabla\cdot\overline{\mathbf{u}}^\ell = 0$, owing to the filtered pressure. In addition, the average amount of strain self-amplification is also held at three times the vorticity stretching for filtered fields,
\begin{equation}
	-\langle \overline{S}_{ij}^\ell \overline{S}_{jk}^\ell \overline{S}_{ki}^\ell \rangle = \frac{3}{4}\langle \overline{\omega}_i^\ell \overline{S}_{ij}^\ell \overline{\omega}_j^\ell \rangle,
	\hspace{0.05\linewidth}
	\text{or}
	\hspace{0.05\linewidth}
	\langle P_{\overline{s}} \rangle = 3 \langle P_{\overline{\omega}} \rangle.
	\label{eq:filtered-Betchov-3}
\end{equation}

\subsection{Energy cascade in terms of filtered velocity gradients}

The local energy cascade rate, $\Pi(\mathbf{x}, t)$, is determined by the filtered strain-rate tensor, the subfilter stress tensor, and how those two align. \citet{Johnson2020} demonstrated how the subfilter stress tensor may be phrased in terms of filtered velocity gradients across all scales $\leq \ell$, so that the energy cascade rate may be written solely in terms of filtered velocity gradients. This result directly connects the restricted Euler singularity with the fact that turbulence generates a net cascade from large to small scales. The derivation proceeds as follows.

 \begin{figure}
	\includegraphics[width=0.49\linewidth]{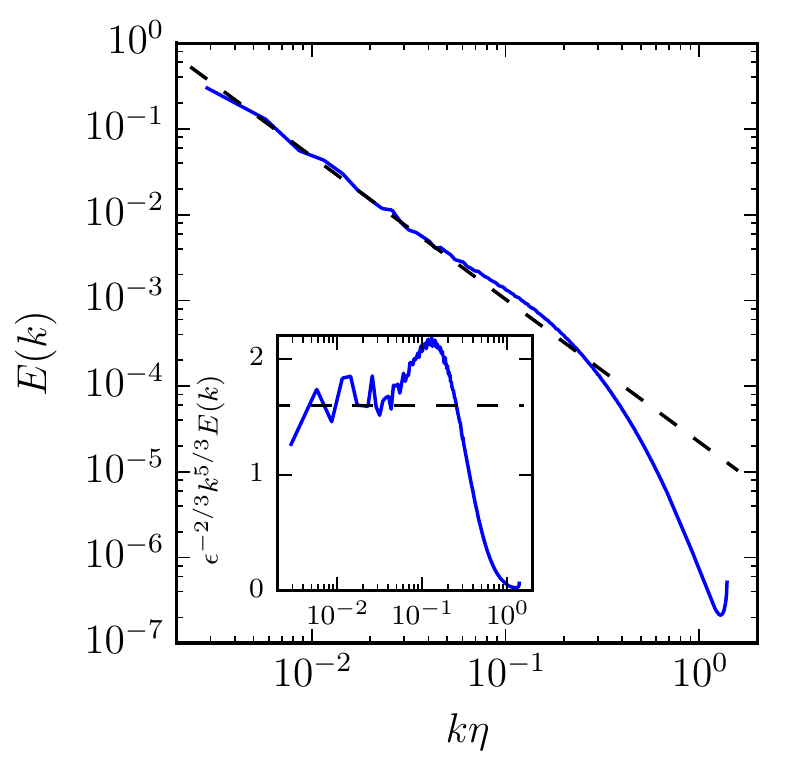}
	\includegraphics[width=0.49\linewidth]{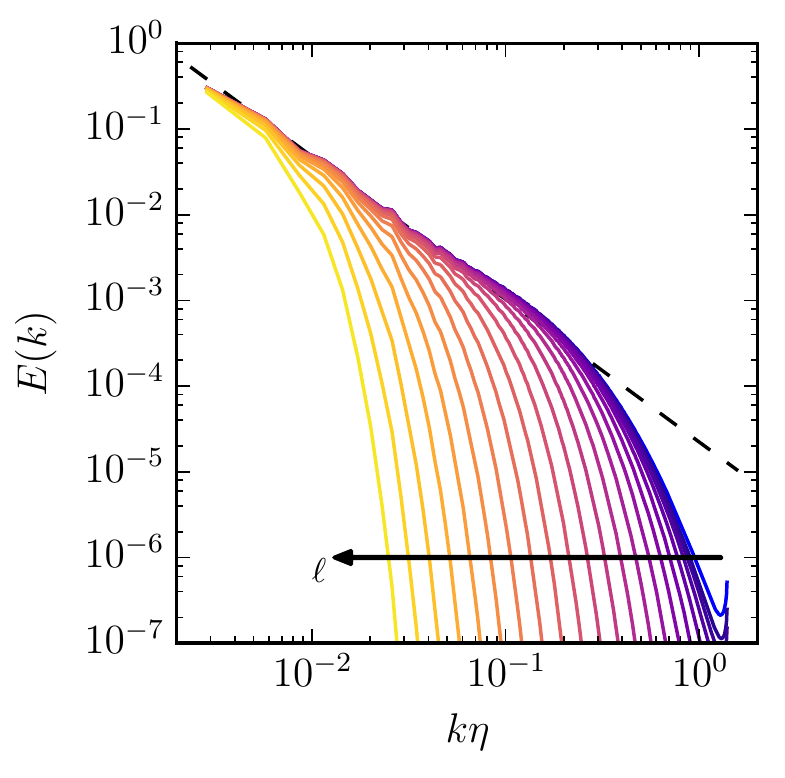}
	\caption{(a) Average energy spectrum from the HIT simulation, and (b) filtered spectra using a Gaussian filter with various values of $0.9 \eta \leq \ell \leq 170 \eta$ spaced evenly in logarithmic space. The black dashed line indicates the inertial range spectrum, $E(k) = 1.6 \epsilon^{2/3} k^{-5/3}$. The inset in panel (a) shows the premultiplied spectrum on a log-linear plot.
	}
	\label{fig:spectrum}
\end{figure}

A Gaussian filter kernel,
\begin{equation}
G_\ell(\mathbf{r}) = \mathcal{N} \exp\left( - \frac{|\mathbf{r}|^2}{2\ell^2}\right),
~~~~~~~
\mathcal{F}\{G_\ell\}(\mathbf{k}) = \exp\left( -\frac{1}{2} |\mathbf{k}|^2 \ell^2 \right),
\label{eq:Gaussian-kernel}
\end{equation}
decays rapidly in both physical space an wavenumber space, and is thus a popular choice for studies involving explicit filters \citep{Borue1998, Domaradski2007b, Eyink2009, Leung2012, Cardesa2015, Lozano2016, Buzzicotti2018, Portwood2020, Vela-Martin2020, Dong2020, Alexakis2020}.
Here, the definition of the filter width, $\ell$, is chosen such that its gradient, $dG/dr$, is maximum at $r=\ell$. That is, velocity increments at separation $r = \ell$ have are the most heavily weighted when constructing the filtered velocity gradient according to Eq. \eqref{eq:filtered-velocity-gradient}. Note that in Figure \ref{fig:inertial_range}a, $2\ell$ appears to correspond to the structure function separation $r$. This happens because, in Eq. \eqref{eq:filtered-velocity-gradient}, the velocity increments with equal and opposite separation vectors of length $\ell$ are essentially combined to form velocity increments at $2\ell$, which are integrated on a half-sphere to form the filtered velocity gradient.

The Gaussian filter has already been demonstrated in physical space in Figures \ref{fig:filter-velocity} and \ref{fig:filter-vorticity}. Figure \ref{fig:spectrum} shows average energy spectra with and without a Gaussian filter. Note that the unfiltered spectrum decays as a (stretched) exponential in the dissipation range \citep{Khurshid2018,Buaria2020}, and the effect of the Gaussian filter moves this exponential-like decay to lower wavenumbers. The unfiltered spectrum shows agreement with the standard Kolmogorov spectrum, $E(k) = 1.6 \epsilon^{2/3} k^{-5/3}$ over a limited range of wavenumbers $k \eta \ll 1$. A `spectral bump', as typically observed, is evident near $0.1 < k\eta < 0.2$ \citep{Mininni2008, Donzis2010}. This is usually associated with a `bottleneck effect' at the end of the cascade in which viscosity is slightly too slow in dissipating energy as it arrives from larger scales, resulting in a slight pile-up of energy. While the filtering approach does smooth out this effect to some degree, it is visible for the filtered strain-rate norm in the inset of Figure \ref{fig:inertial_range}(b) in the range $5 < \ell/\eta < 10$.

Because the Gaussian kernel serves as the Green's function for the diffusion equation, a filtered quantity is the solution to
\begin{equation}
\frac{\partial \overline{a}^\ell}{\partial (\ell^2)} = \frac{1}{2} \nabla^2 \overline{a}^\ell,
\hspace{0.05\linewidth}
\text{with initial condition}
\hspace{0.05\linewidth}
\overline{a}^{\ell=0}(\mathbf{x}) = a(\mathbf{x})
\label{eq:filter-diffusion-equation}
\end{equation}
where the square of the filter width, $\ell^2$, serves as a time-like variable. The initial condition at $\ell = 0$ is the unfiltered field. From this observation, it is straightforward to show that any generalized second moment may be written as
\begin{equation}
\frac{\partial \tau_\ell(a, b)}{\partial (\ell^2)} = \frac{1}{2} \nabla^2 \tau_{\ell}(a, b) + \frac{\partial \overline{a}^\ell}{\partial x_k} \frac{\partial \overline{b}^\ell}{\partial x_k},
\hspace{0.02\linewidth}
\text{with initial condition}
\hspace{0.02\linewidth}
\tau_{\ell=0}(a, b) = 0.
\label{eq:moment-diffusion-equation}
\end{equation}
Thus, the generalized second moment is the solution to a forced diffusion equation with zero initial condition. The forcing is the product of filtered gradients as a function of scale (the time-like variable). The formal solution for this forced diffusion equation is
\begin{equation}
\tau_\ell(a,b) = \int_{0}^{\ell^2} d\alpha ~
\overline{
	\overline{\frac{\partial a}{\partial x_k}}^{\sqrt{\alpha}} \overline{\frac{\partial b}{\partial x_k}}^{\sqrt{\alpha}}
}^{\beta},
\label{eq:moment-diffusion-solution}
\end{equation}
where $\beta \equiv \sqrt{\ell^2 - \alpha}$ is the conjugate filter such that successively filtering at $\sqrt{\alpha}$ and $\beta$ is equivalent to applying a single filter at scale $\ell$,
\begin{equation}
\overline{\overline{a}^{\sqrt{\alpha}}}^{\beta} = \overline{a}^{\ell}.
\label{eq:successive-filter-relation}
\end{equation}
The formal solution in Eq. \eqref{eq:moment-diffusion-equation} contains two components,
\begin{equation}
\tau_\ell(a,b) = \ell^2 \overline{\frac{\partial a}{\partial x_i}}^{\ell} \overline{\frac{\partial b}{\partial x_i}}^{\ell}
+ \int_{0}^{\ell^2} d\alpha ~
\tau_\beta\left( \overline{\frac{\partial a}{\partial x_k}}^{\sqrt{\alpha}}, \overline{\frac{\partial b}{\partial x_k}}^{\sqrt{\alpha}} \right).
\label{eq:Gaussian-relation}
\end{equation}
The first component is the product of gradients at the filter scale $\ell$ and the second component involves subfilter scale generalized second moments of gradients filtered at scales smaller than $\ell$.
Equation \eqref{eq:Gaussian-relation} is applied to the subfilter stress, $\tau_\ell(u_i, u_j)$, by setting $a = u_i$ and $b = u_j$.  Contracting with $\overline{S}_{ij}^\ell$ to compute the energy cascade rate, $\Pi^\ell$, results in
\begin{equation}
\Pi^\ell = \Pi_{s1}^\ell + \Pi_{\omega 1}^\ell + \Pi_{s2}^\ell + \Pi_{\omega 2}^\ell + \Pi_{c}^\ell.
\label{eq:Pi-decomposition}
\end{equation}
The five components of Eq. \eqref{eq:Pi-decomposition} are defined and interpreted as follows.

The first two terms involve velocity gradients filtered at scale $\ell$ only. First, the self-amplification of the strain-rate at scale $\ell$ transfer energy across scale $\ell$ at a rate,
\begin{equation}
\Pi_{s1}^\ell = \ell^2 P_{\overline{s}} = -\ell^2 \overline{S}_{ij}^\ell  \overline{S}_{jk}^\ell  \overline{S}_{ki}^\ell
= -3 \ell^2 \lambda_1^\ell \lambda_2^\ell \lambda_3^\ell,
\label{eq:resolved-strain-amp}
\end{equation}
where where $\lambda_i^\ell$ are the $3$ eigenvalues of the strain-rate tensor filtered at scale $\ell$. Similarly, the vorticity at scale $\ell$ is stretched by the strain-rate at the same scale, which transfers energy
\begin{equation}
\Pi_{\omega 1}^\ell = \ell^2 P_{\overline{\omega}} 
= \frac{1}{4} \ell^2 \overline{S}_{ij}^\ell \overline{\omega}_{i}^\ell \overline{\omega}_j^\ell
= \frac{1}{4} \ell^2 \left|\overline{\boldsymbol{\omega}}^\ell\right|^2 \sum_{i=1}^{3} \lambda_i^\ell \cos^2(\theta_{\omega,i}^\ell),
\label{eq:resolved-vorticity-stretch}
\end{equation}
 where $\theta_{\omega,i}^\ell$ are the angles between each corresponding eigenvector and the vorticity vector filtered at scale $\ell$. A subscript ``1'' is used to denote that these rates involve quantities at a single filter scale. In fact, these two terms appear both in Eq. \eqref{eq:Pi-decomposition} and in the velocity gradient evolution, Eq. \eqref{eq:filtered-restricted-Euler-norm}.
The same processes responsible for increasing the filtered velocity gradient magnitude also redistribute energy to sub-filter scales, thus connecting the restricted Euler singularity with the energy cascade. It is also possible to obtain these two as leading order terms in an infinite expansion in at least two ways.

First, using spatial filtering with a more general filter shape, the Taylor expansion of the Leonard stress is led by these two terms \citep{Pope2000}. Truncation of this expansion is sometimes called the Clark model, or tensor diffusivity model \citep{Clark1979, Borue1998}. This model performs well in \textit{a priori} tests, but does not remove large scale energy at a sufficient rate when used with large-eddy simulations \citep{Vreman1996, Vreman1997}. Intriguingly, it may be shown that the Lagrangian memory effects in the evolution of the subfilter stress tensor are correctly mimicked by the tensor diffusivity model \citep{Johnson2020b}. In practice, the tensor diffusivity model is often supplemented with an eddy viscosity model \citep{Clark1979, Vreman1996, Vreman1997}. \citet{Eyink2006} extended this result to a multi-scale gradient expansion for the full subfilter stress tensor, but still relied on truncating infinite series.

\citet{Carbone2020} provided an alternative derivation of single-scale vorticity stretching and strain self-amplification as leading order terms in an infinite series. In that case, the K\'{a}rm\'{a}n-Howarth-Monin equation \citep{Karman1938, Monin1975, Hill2001} is used with filtered velocity gradients substituted as the leading order term in the evaluation of velocity increments.  Importantly, the result highlights agreement between velocity increment and filtering approaches to the energy cascade, indicating a degree of objectivity to this result (see, e.g., objections in \citet{Tsinober2009}).

The true advantage of the result from \citet{Johnson2020}, Eq. \eqref{eq:Pi-decomposition}, lies in the remaining three terms. These cascade rates involve interactions between the strain-rate filtered at scale $\ell$ and velocity gradients at scales smaller than $\ell$. They are given a subscript `2' to denote their multiscale nature (specifically, the sum of interactions involving two different scales). First, the cascade rate due to the amplification of small-scale strain by larger-scale strain is
\begin{equation}
\Pi_{s2}^\ell = - \overline{S}_{ij}^\ell \int_{0}^{\ell^2} d\alpha~ \tau_{\beta}\left( \overline{S}_{jk}^{\sqrt{\alpha}}, \overline{S}_{ki}^{\sqrt{\alpha}} \right)
= - \int_{0}^{\ell^2} d\alpha~ \overline{S}_{ij}^\ell \tau_{\beta}\left( \overline{S}_{jk}^{\sqrt{\alpha}}, \overline{S}_{ki}^{\sqrt{\alpha}} \right).
\label{eq:multiscale-strain-amp}
\end{equation}
Similarly, the stretching of small-scale vorticity by larger-scale strain is
\begin{equation}
\Pi_{\omega 2}^\ell = \overline{S}_{ij}^\ell \int_{0}^{\ell^2} d\alpha~ \tau_\beta\left(\overline{\omega}_i^{\sqrt{\alpha}}, \overline{\omega}_j^{\sqrt{\alpha}}\right)
= \int_{0}^{\ell^2} d\alpha~\overline{S}_{ij}^\ell \tau_\beta\left(\overline{\omega}_i^{\sqrt{\alpha}}, \overline{\omega}_j^{\sqrt{\alpha}}\right).
\label{eq:multiscale-vorticity-stretch}
\end{equation}
These two multiscale terms are analogous in physical interpretation to the single-scale terms discussed earlier. They underscore the importance of multiscale interactions in the cascade and, as will be seen, provide a view into the relative scale-locality of the energy cascade.

Finally, the last term in Eq. \eqref{eq:Pi-decomposition} represents the contribution to the energy cascade rate from larger-scale strain-rates acting on smaller-scale strain-vorticity covariance,
\begin{equation}
\Pi_{c}^\ell = \Pi_{c2}^\ell = \overline{S}_{ij}^\ell \int_{0}^{\ell^2} d\alpha~ \left[ \tau_\beta\left( \overline{S}_{jk}^{\sqrt{\alpha}}, \overline{\Omega}_{ki}^{\sqrt{\alpha}} \right)
- \tau_\beta\left( \overline{\Omega}_{jk}^{\sqrt{\alpha}}, \overline{S}_{ki}^{\sqrt{\alpha}} \right) \right]
\label{eq:multiscale-cross}
\end{equation}
This term is given the subscript `c', denoting a cross term between both strain-rate and vorticity. It is the only of the five terms not interpretable as vorticity stretching of strain self-amplification. Its meaning is less obvious, but one potential interpretation is given in \S \ref{sec:2D} while considering the inverse energy cascade in two dimensions.

\begin{figure}
	\includegraphics[width=0.5\linewidth]{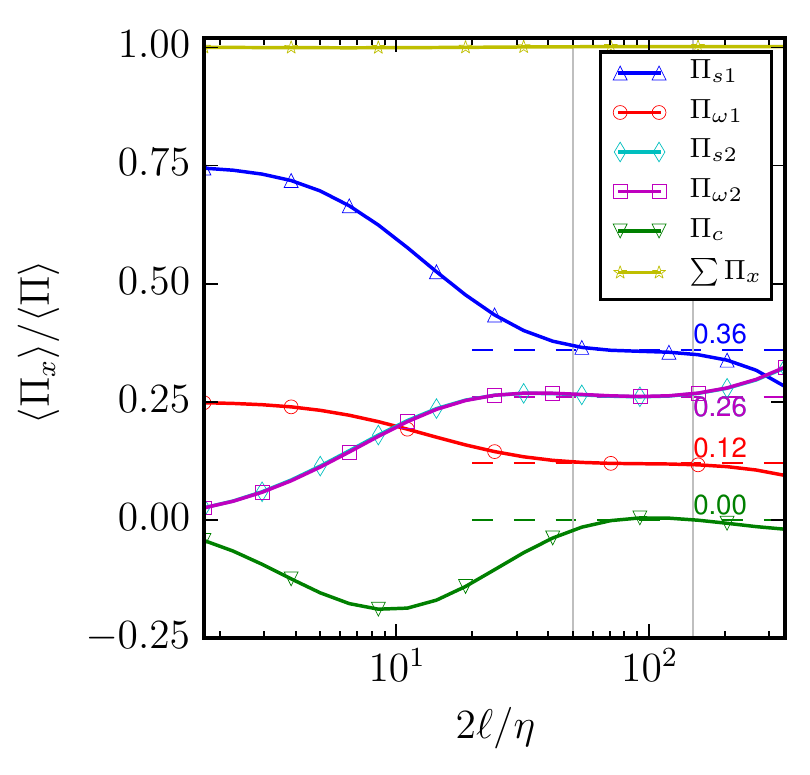}
	\includegraphics[width=0.495\linewidth]{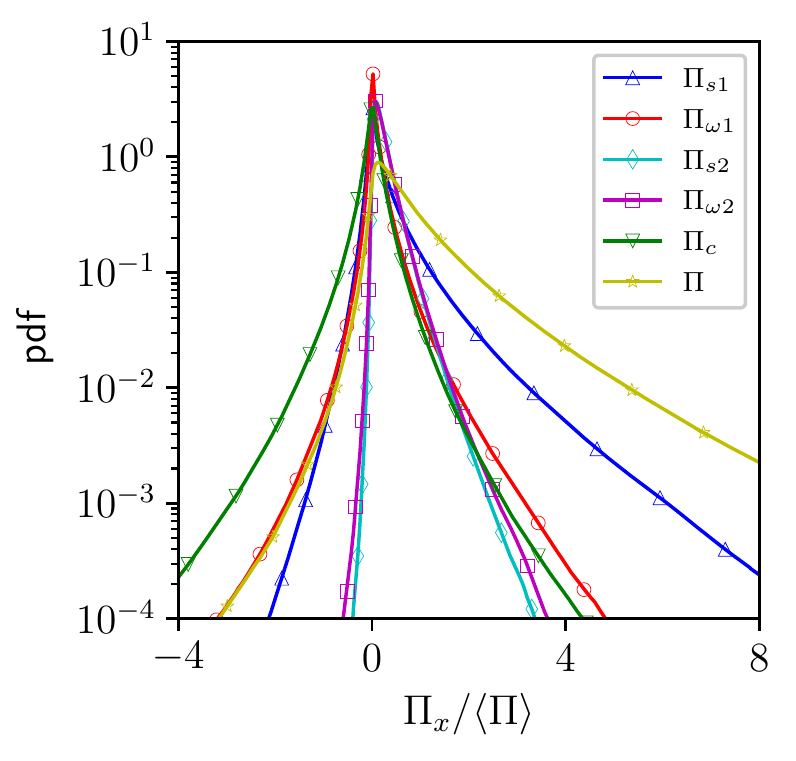}
	\caption{(a) The mean contribution of each cascade mechanism as a function of filter width. (b) The PDFs of each component of the energy cascade rate from Eq. \eqref{eq:Pi-decomposition} at $\ell = 46 \eta$.
		The two vertical gray lines are at $2\ell/\eta = 50$ and $150$, indicating the approximate inertial range of scales. The integral length scale is at $2 L / \eta = 920$.
	}
	\label{fig:Pi}
\end{figure}

Figure \ref{fig:Pi}(a) shows the net contribution of each dynamical mechanism to the cascade rate, as a fraction of the total cascade rate, which is shown Figure \ref{fig:inertial_range}a. First, the sum of all five contributions is equal to unity, confirming the correctness of Eq. \eqref{eq:Pi-decomposition}. The point-wise accuracy of Eq. \eqref{eq:Pi-decomposition} is also confirmed in Appendix \ref{sec:app}, which also includes a brief study of its accuracy when some terms are neglected.

In the dissipative range, $\ell \sim \eta$, the three multiscale terms are small and the two single-scale terms comprise the cascade at the specified ratio of $3$:$1$. In the range of filter widths previously identified as the inertial range, $25 < \ell/\eta < 75$ for this simulation, the fractional contribution of each cascade rate is relative constant, indicating some degree of self-similarity. Each plateau is fitted to two decimal places as reported in the figure. Note that $\langle \Pi_c \rangle \approx 0$ so that the other four terms are responsible for establishing the cascade rate. The sum of the two single-scale rates account for approximately half of the full cascade rate, with their multiscale analogs accounting for the other half. The amplification of strain-rates, both single-scale and multi-scale, account for roughly $5/8$ of the energy cascade rate, with the remaining $3/8$ due to vorticity stretching. Interestingly, the strain-vorticity covariance term supplies a negative cascade rate at filter widths corresponding to the spectral bump, offering a potential clue to the dynamical causes of the bottleneck effect.

The probability density functions (PDF) of each contribution to the energy cascade rate from Eq. \eqref{eq:Pi-decomposition} are shown in Figure \ref{fig:Pi}(b) with filter width in the middle of the inertial range. The total cascade rate PDF is strongly skewed with relatively rare backscatter events. The net backscatter is less than $2\%$ of the total average cascade rate. The single-scale strain self-amplification, $\Pi_{s1}$, is also strongly skewed and has the most probable extreme positive events of any of the five components. The single-scale vorticity stretching, $\Pi_{\omega 1}$ is much less skewed toward downscale energy cascade. The multiscale and strain amplification and vorticity stretching, $\Pi_{s2}$ and $\Pi_{\omega 2}$, are extremely skewed and almost never negative. Finally, the amplification of subfilter strain-vorticity covariance, $\Pi_c$, is relatively symmetric and has the most probable extreme negative events.

\subsection{Implications for the inverse cascade in 2D turbulence}\label{sec:2D}

The main topic of this paper is the energy cascade in three-dimensional flows. However, it is worthwhile to comment on implications for the inverse cascade of energy in two dimensions. Of the five terms on the right side of Eq. \eqref{eq:Pi-decomposition}, only the fifth term is non-zero. All vorticity stretching and strain amplification terms vanish exactly in 2D incompressible flows. Thus,
\begin{equation}
	\Pi^\ell = \Pi_{c2}^\ell = \overline{S}_{ij}^\ell \int_{0}^{\ell^2} d\alpha~ \left[ \tau_\beta\left( \overline{S}_{jk}^{\sqrt{\alpha}}, \overline{\Omega}_{ki}^{\sqrt{\alpha}} \right)
	- \tau_\beta\left( \overline{\Omega}_{jk}^{\sqrt{\alpha}}, \overline{S}_{ki}^{\sqrt{\alpha}} \right) \right].
	\label{eq:Pi_2D}
\end{equation}
Taking the eigenframe of the strain-rate at scale $\ell$,
\begin{equation}
	\Pi^\ell(\mathbf{x}) = 2 \int_{0}^{\ell^2} d\alpha \iiint d\mathbf{r}~ G_\beta(\mathbf{r})~ \lambda^\ell(\mathbf{x})~ \lambda^{\sqrt{\alpha}}(\mathbf{x}+\mathbf{r})~ \overline{\omega}^{\sqrt{\alpha}}(\mathbf{x}+\mathbf{r})~ \sin 2 \phi(\mathbf{x},\mathbf{r}),
\end{equation}
where $\phi(\mathbf{x},\mathbf{r})$ is the angle of the eigenvectors of $\overline{\mathbf{S}}^{\sqrt{\alpha}}(\mathbf{x}+\mathbf{r})$ with respect to the eigenvectors of $\overline{\mathbf{S}}^\ell(\mathbf{x})$. Here, the eigenvalues of the strain rate are $\lambda$ and $-\lambda$. If the strain-rate eigenvectors at $\ell$ and $\sqrt{\alpha}$ align parallel or perpendicular, then there is no transfer of energy. Maximum transfer of energy across scale $\ell$ occurs for $\pm 45^\text{o}$, with the direction of the transfer also depending on the sign of vorticity. An inverse cascade is supported when smaller-scale strain-rate is misaligned with larger-scale strain-rate in the opposite rotational direction as the vorticity.

\begin{figure}
	\centering
	\includegraphics[page=4, width=0.99\linewidth]{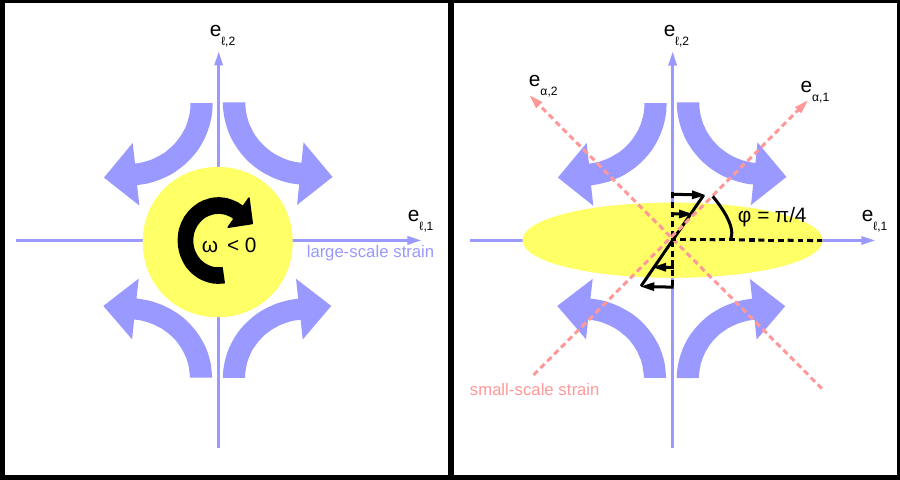}
	\caption{Simplified schematic of the vortex thinning mechanism driving an inverse energy cascade in two-dimensional turbulence.}
	\label{fig:2D-sketch}
\end{figure}

One such (cartoon-ish) scenario is sketched in Figure \ref{fig:2D-sketch}. As a clockwise vortex ($\omega < 0$) is subjected to a larger-scale strain-rate, it is flattened or thinned into a shear layer with strain-rate eigenvalues at $45^\text{o}$ from the larger-scale strain. The resulting alignment produces a maximum inverse cascade rate. This is the vortex thinning mechanism of the 2D inverse cascade \citep{Kraichnan1976,Chen2006,Xiao2009}.

Therefore, Equation \eqref{eq:Pi-decomposition} represents a comprehensive view of inter-scale energy transfer in two- and three-dimensional turbulence. It is interesting to note (G. L. Eyink, private communication) that this approach may be thought of as a sort of differential renormalization group analysis \citep{Eyink2018}. While vorticity stretching and (to some degree) strain self-amplification are commonly discussed as cascade mechanisms in three-dimensional turbulence, it should not be a surprise that the vortex thinning mechanism posited for the 2D inverse cascade is at least a possible mechanism for energy transfer in 3D as well. Thus, while Eq. \eqref{eq:Pi-decomposition} from \citet{Johnson2020} clearly identifies the roles of vorticity stretching and strain self-amplification, it also demonstrates how vortex thinning could be active, even if that activity may be negligible outside a particular range of scales in practice.

In two dimensions $\Pi_{c2}$ represents the inverse cascade. In the inertial range for three-dimensional turbulence, this vortex thinning term is evidently small compared to vorticity stretching and strain self-amplification, both of which drive a forward cascade to small scales. However, Figure \ref{fig:Pi}a also reveals that the vortex thinning term does provide a backscatter contribution for scales in between the viscous and inertial ranges, suggesting that the vortex thinning mechanism as a potential dynamical cause of the bottleneck effect.

\section{Efficiency of the energy cascade}\label{sec:efficiency}

\citet{Ballouz2018} defined an energy cascade efficiency as
\begin{equation}
\Gamma_{BO}^\ell
= \frac{\Pi^\ell}{\Pi_{\max,BO}^\ell},
\hspace{0.04\linewidth}
\text{where}
\hspace{0.04\linewidth}
\Pi_{\max,BO}^\ell = \left\lbrace
\begin{array}{c c}
	\lambda_1^\ell \mu_1^\ell + \lambda_2^\ell \mu_2^\ell + \lambda_3^\ell \mu_3^\ell & \Pi^\ell > 0 \\
	- \left(\lambda_1^\ell \mu_3^\ell + \lambda_2^\ell \mu_2^\ell + \lambda_3^\ell \mu_1^\ell \right) & \Pi^\ell < 0
\end{array}
\right.
\label{eq:efficiency-Ballouz}
\end{equation}
is the maximum energy transfer possible for a fixed set of eigenvalues, $\mu_i$ and $\lambda_j$, for $\mathring{\boldsymbol{\tau}}^\ell$ and $\overline{\mathbf{S}}^\ell$, respectively. Perfect alignment between eigenvectors of the $\mathring{\boldsymbol{\tau}}^\ell$ and $\overline{\mathbf{S}}^\ell$ is required for an efficiency of $\Gamma^\ell = 1$. Under this definition, the average cascade efficiency is 
approximately $40\%$ for the top-hat filter for filter widths much larger than the Kolmogorov scale \citep{Ballouz2020b}. 

Here, a slightly different definition of efficiency is used,
\begin{equation}
\Gamma^\ell = \frac{\Pi^\ell}{\Pi_{\max}^\ell}
= \frac{-\mathring{\boldsymbol{\tau}}^\ell : \overline{\mathbf{S}}^\ell}{\| \mathring{\boldsymbol{\tau}}^\ell\| \|\overline{\mathbf{S}}^\ell \|}
= \frac{\sum_{i=1}^{3} \sum_{j=1}^{3} \lambda_i^\ell \mu_j^\ell \cos^2(\theta_{ij}^\ell)}{\left( {\lambda_1^\ell}^2 + {\lambda_2^\ell}^2 + {\lambda_3^\ell}^2 \right)^{1/2} \left( {\mu_1^\ell}^2 + {\mu_2^\ell}^2 + {\mu_3^\ell}^2 \right)^{1/2}}.
\label{eq:efficiency}
\end{equation}
In addition to the re-alignment of eigenvectors, this definition of $\Pi_{\max}^\ell$ in the denominator allows for the ratio of eigenvalues to be rearranged at fixed Frobenius norm to further maximize the flux. Thus, unity efficiency $\Gamma^\ell = 1$ requires not only perfect alignment of (respective) eigenvectors, $\cos^2(\theta_{ij}) = \delta_{ij}$, but also the proportionality of eigenvalues,
\begin{equation}
\frac{\lambda_1}{\mu_1} = \frac{\lambda_2}{\mu_2} = \frac{\lambda_3}{\mu_3}.
\end{equation}
If follows that
\begin{equation}
\Pi_{\max}^\ell \geq \Pi_{BO\max}^\ell,
~~~~~~~
\text{therefore}
~~~~~~~
\Gamma^\ell \leq \Gamma_{BO}^\ell.
\end{equation}
It is important that $\mathring{\boldsymbol{\tau}}$ is the deviatoric part of the sub-filter stress when constructing the denominator of Eq. \eqref{eq:efficiency}. This is significant because the inclusion of the isotropic part of the sub-filter stress would increase its norm without increasing the maximum potential value of its contraction with the deviatoric strain-rate tensor. 

Phrasing $\Pi^\ell$ in terms of vorticity stretching and strain self-amplification, Eqs. \eqref{eq:Pi-decomposition}-\eqref{eq:multiscale-cross}, allows for defining the efficiency of each dynamical mechanism in transferring energy downscale. Each component of $\Pi^\ell$ can be written as an inner product of $\overline{\mathbf{S}}^\ell$ with a component of the subfilter stress tensor.
For example, the deviatoric part of the sub-filter stress due to single-scale strain self-amplification is
\begin{equation}
\left(\mathring{\boldsymbol{\tau}}_{s1}^\ell\right)_{ij} = \ell^2 \left( \overline{S}_{ik}^\ell \overline{S}_{jk}^\ell - \frac{1}{3} \|\overline{\mathbf{S}}^\ell\|^2 \delta_{ij} \right).
\end{equation}
The efficiency of the strain self-amplification, $\Pi_{s1}$, in the downscale transfer of energy is
\begin{equation}
\Gamma_{s1}^\ell 
= \frac{-\mathring{\boldsymbol{\tau}}_{s1}^\ell:\overline{\mathbf{S}}^\ell}{\| \mathring{\boldsymbol{\tau}}_{s1}^\ell \|\| \overline{\mathbf{S}}^\ell \|}
= \frac{\sqrt{6} ~\Pi_{s1}^\ell}{\ell^2 \| \overline{\mathbf{S}}^\ell \|^3}
= \frac{-\sqrt{6} ~\overline{S}_{ij}^\ell\overline{S}_{jk}^\ell\overline{S}_{ki}^\ell}{\| \overline{\mathbf{S}}^\ell \|^3}
= \frac{- 3 \sqrt{6} ~\lambda_1^\ell \lambda_2^\ell \lambda_3^\ell}{\left( {\lambda_1^\ell}^2 + {\lambda_2^\ell}^2 + {\lambda_3^\ell}^2 \right)^{3/2}}.
\label{eq:efficiency-resolved-strain-amp}
\end{equation}
which the reader may recognize as the $s^*$ parameter of \citet{Lund1994} applied to the filtered strain-rate tensor. Thus, the strong bias of turbulence toward $\lambda_{2}^\ell > 0$ manifests as a bias toward positive efficiency of downscale energy flux. In fact, the PDF of $s^*$ has been consistently shown to peak at $s^* = 1$ ($\Gamma_{s1} = 1$), i.e., $\lambda_{1}^\ell = \lambda_{2}^\ell = -\frac{1}{2}\lambda_{3}^\ell$. When this is true, there is one eigenvector along which the flow is strongly squeezing, with weaker stretching occurring along perpendicular direction. In the opposite case, $\Gamma_{s1}^\ell = -1$, there is strong stretching along one eigenvector and weaker squeezing along perpendicular directions, i.e. $\frac{1}{2}\lambda_1^\ell = - \lambda_2^\ell = -\lambda_3^\ell$. This situation corresponds to maximum efficiency in reverse (upscale) energy transfer and is relatively rare in turbulent flows.

Note that use of the Ballouz-Ouellette efficiency, i.e., not allowing for re-arranging of eigenvalue ratios when constructing the maximum potential flux, would result in a unity efficiency for the resolved strain self-amplification. The reason for this is that $\mathring{\boldsymbol{\tau}}_{s1}^\ell$ is, by definition, always perfectly aligned with $\overline{\mathbf{S}}^\ell$ so that the cascade rate by this mechanism is only a function of the eigenvalues. This consideration was part of the reasoning motivating the definition of efficiency used here.

Another reason to use Eq. \eqref{eq:efficiency} rather than \eqref{eq:efficiency-Ballouz} is that the former provides a stricter test of whether the subfilter stress conforms to the eddy viscosity hypothesis, $\mathring{\tau}_{ij}^\ell = - 2 \nu_t \overline{S}_{ij}^\ell$. The eddy viscosity ansatz asserts that the subfilter stress not only aligns its eigenvectors with those of the filtered strain-rate tensor, but also that their eigenvalues are proportional. Thus, with the present definition of efficiency,
\begin{equation}
	\Gamma^\ell = \left\lbrace
	\begin{array}{l l}
		1, & \text{eddy viscosity (downscale cascade)} \\
		0, & \text{zero cascade rate} \\
		-1, & \text{negative eddy viscosity (inverse cascade)}
	\end{array}
	\right.,
\end{equation}
which can be applied individually to each term in Eq. \eqref{eq:Pi-decomposition} to test the extent to which each cascade mechanism may be accurately modeled by an eddy viscosity.

Unlike the strain self-amplification, vorticity stretching suffers from eigenvector misalignment in addition to non-optimal eigenvalue ratios. The efficiency of the energy cascade due to single-scale vorticity stretching is
\begin{equation}
\Gamma_{\omega 1}^\ell 
= \frac{-\mathring{\boldsymbol{\tau}}_{\omega 1}^\ell:\overline{\mathbf{S}}^\ell}{\| \mathring{\boldsymbol{\tau}}_{\omega 1}^\ell \|\| \overline{\mathbf{S}}^\ell \|}
= \frac{\sqrt{6}~ \Pi_{\omega 1}^\ell}{\ell^2 \|\overline{\boldsymbol{\Omega}}^\ell \|^2 \| \overline{\mathbf{S}}^\ell \|}
= \frac{\sqrt{6}~ \overline{\omega}_i^\ell \overline{\omega}_j^\ell \overline{S}_{ij}^\ell }{2 |\overline{\boldsymbol{\omega}}^\ell |^2 \| \overline{\mathbf{S}}^\ell \|}
= \frac{\sqrt{6}~ \sum_{i=1}^{3}\lambda_i^\ell \cos^2\left(\theta_{\omega,i}^\ell\right) }{2 \left( {\lambda_1^\ell}^2 + {\lambda_2^\ell}^2 + {\lambda_3^\ell}^2 \right)^{1/2}},
\label{eq:efficiency-resolved-vorticity-stretch}
\end{equation}
where
\begin{equation}
\left(\mathring{\boldsymbol{\tau}}_{\omega 1}^\ell\right)_{ij}
= \ell^2 \left( \overline{\Omega}_{ik}^\ell \overline{\Omega}_{jk}^\ell - \frac{1}{3} \|\overline{\mathbf{\Omega}}^\ell\|^2 \delta_{ij} \right)
= -\frac{1}{4}\ell^2 \left( \overline{\omega}_i^\ell \overline{\omega}_j^\ell - \frac{1}{3} |\overline{\boldsymbol{\omega}}^\ell|^2 \delta_{ij} \right).
\end{equation}
This efficiency is maximum, $\Gamma_{\omega 1}^\ell = 1$, when two conditions are satisfied. First, the vorticity is perfectly aligned with the maximum eigenvalue, $\cos(\theta_{\omega,i}^\ell) = \delta_{i1}$. Second, the strain-rate tensor must have the configuration  $\frac{1}{2}\lambda_1^\ell = - \lambda_2 = -\lambda_3$ so that the magnitude of the stretching is maximized. Note that the condition for maximum vorticity stretching efficiency is that of $\Gamma_{s1}^\ell = -1$. When $\Gamma_{s1} = 1$, the maximum efficiency of vorticity stretching is limited, $-1 \leq \Gamma_{\omega 1} \leq 0.5$. Therefore, the chaos of turbulent dynamics aside, the efficiencies of strain-rate self-amplification and vorticity stretching cannot be simultaneously maximum. In fact, the asymptotic behavior of solutions to the restricted Euler equation as they approach the finite-time singularity is $\Gamma_{s1}^\ell = 1$ and $\Gamma_{\omega 1}^\ell = 0.5$.

The efficiencies for the multiscale terms are similarly defined and interpreted,
\begin{equation}
	\Gamma_{s2}^\ell = \frac{-\mathring{\boldsymbol{\tau}}_{s2}^\ell:\overline{\mathbf{S}}^\ell}{\| \mathring{\boldsymbol{\tau}}_{s2}^\ell \|\| \overline{\mathbf{S}}^\ell \|},
	\hspace{0.05\linewidth}
	\Gamma_{\omega 2} = \frac{-\mathring{\boldsymbol{\tau}}_{\omega 2}^\ell:\overline{\mathbf{S}}^\ell}{\| \mathring{\boldsymbol{\tau}}_{\omega 2}^\ell \|\| \overline{\mathbf{S}}^\ell \|},
	\hspace{0.05\linewidth}
	\Gamma_{c}^\ell = \frac{-\mathring{\boldsymbol{\tau}}_{c}^\ell:\overline{\mathbf{S}}^\ell}{\| \mathring{\boldsymbol{\tau}}_{c}^\ell \|\| \overline{\mathbf{S}}^\ell \|}.
\end{equation}
In terms of filtered strain-rate eigenvalues, $\lambda_i^\ell$, subfilter stress eigenvalues, $\mu_{x,j}$, and eigenvector alignment angles, $\theta_{x,ij}$, the efficiencies of the multiscale cascade rates are
\begin{equation}
\Gamma_{x}^\ell
= \frac{\sum_{i=1}^{3} \sum_{j=1}^{3} \lambda_i^\ell \mu_{x,j}^\ell \cos^2(\theta_{x,ij}^\ell)}{\left( {\lambda_1^\ell}^2 + {\lambda_2^\ell}^2 + {\lambda_3^\ell}^2 \right)^{1/2} \left( {\mu_{x,1}^\ell}^2 + {\mu_{x,2}^\ell}^2 + {\mu_{x,3}^\ell}^2 \right)^{1/2}},
\hspace{0.03\linewidth}
\text{where}
\hspace{0.03\linewidth}
x = s2, \omega 2, c.
\label{eq:efficiency-multiscale-strain-amp}
\end{equation}

\begin{figure}
	\includegraphics[width=0.49\linewidth]{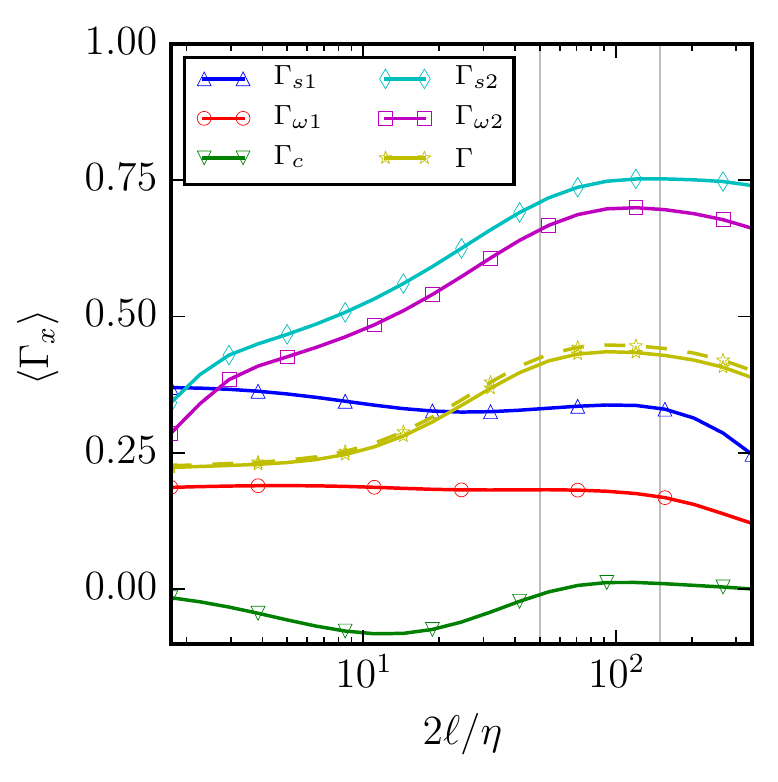}
	\includegraphics[width=0.495\linewidth]{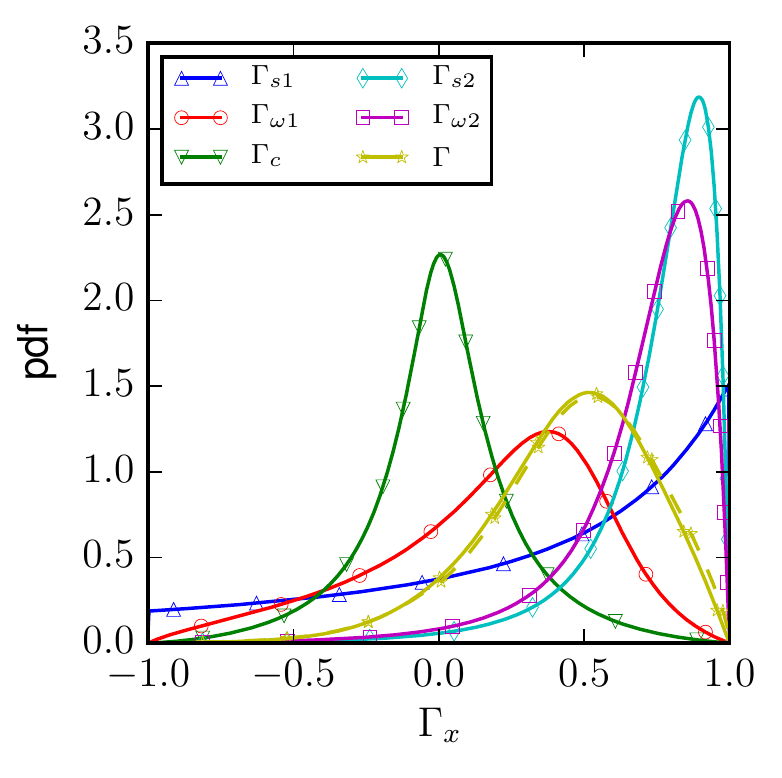}
	\caption{Cascade efficiencies for each term in Eq. \eqref{eq:Pi-decomposition}: (a) average efficiency as a function of filter width, (b) PDF of efficiency at $\ell = 46\eta$. The dashed line shows the Ballouz-Ouellette efficiency.}
	\label{fig:gama}
\end{figure}

The average cascade efficiencies from DNS of HIT is shown in Figure \ref{fig:gama}a as a function of filter width. At a filter width chosen in the middle of the inertial range, $\ell = 46\eta$, the PDF of each term is also shown in Figure \ref{fig:gama}b. The overall cascade efficiency is less than $25\%$ in the viscous range and grows above $40\%$ in the inertial range. The overall efficiency based the Ballouz-Ouellette definition is only slightly larger than the present one. The PDF of the total cascade rate peaks near 50\%, and negative efficiencies below $-50\%$ extremely rare. Indeed, the eddy viscosity approximation for subfilter stresses lacks a high degree physical fidelity in general.

The single-scale terms, $\Pi_{s1}$ and $\Pi_{\omega 1}$, are less efficient than the total subfilter activity in supporting an downscale cascade in the inertial range. In fact, their efficiency is relatively constant as a function of scale from the viscous range through the inertial range. The strain self-amplification efficiency PDF peaks at $\Gamma_{s1}^\ell = 1$, but also allows for substantial negative efficiencies. Meanwhile, the most commonly observed efficiency for single-scale vorticity stretching is slightly less than $50\%$, but negative efficiencies for this mechanism are also not particularly rare. Thus, the signature of restricted Euler dynamics is observed in the location of the efficiency PDF peaks, namely, at $\Gamma_{s1}^\ell = 1$ and $\Gamma_{\omega 1}^\ell = 0.5$. Nevertheless, the mean efficiency is quite low due to the fluctuations.

In constrast, the three multiscale processes have more sharply peaked efficiency PDFs. Multiscale vorticity stretching and strain self-amplification show a strong tendency toward eddy viscosity-like behaviors, with mean efficiencies near $70\%$ and $75\%$, respectively, in the inertial range. Indeed these terms rarely generate backscatter, and when they do, it is inefficient in alignment. While the restricted Euler dynamics may be invoked for the single-scale terms, the smaller-scale strain-rates and vorticities comprising $\mathring{\boldsymbol{\tau}}_{s2}^\ell$ and $\mathring{\boldsymbol{\tau}}_{\omega 2}^\ell$ evolve on faster timescales compared with $\overline{\mathbf{S}}^\ell$, coming closer to satisfying conditions of scale-separation that leads to eddy viscosity physics. It is known, for example, that finer-scale vorticity aligns more readily with the strongest extensional direction of a larger-scale strain-rate \citep{Leung2012, Fiscaletti2016}, which is physically related to the time-lag in alignment elucidated by \citet{Xu2011}. The Lagrangian behavior of these terms deserves further consideration in future work.

The PDF of the strain-vorticity covariance term, $\Pi_c$, is also strongly peaked. Unlike the vorticity stretching and strain self-amplification terms, the most common value of $\Pi_c$ is zero. The PDF is symmetric and the mean value is very small throughout the inertial range. While it is does not appear correlated with $\overline{\mathbf{S}}^\ell$, it is not known whether $\mathring{\boldsymbol{\tau}}_{c2}$ is strongly correlated with other quantities filtered at scale $\ell$ to facilitate LES modeling of this term.

\section{Scale-locality of the energy cascade}\label{sec:locality}

It has been presumed in the discussion so far that the inter-scale energy transfer in turbluence may be faithfully characterized as a cascade. However, it is useful to leverage the present developments to examine the scale-locality of energy transfer. A distinct advantage of Eq. \eqref{eq:Pi-decomposition} is that its formulation allows this question to be addressed in the context of vorticity stretching and strain self-amplification mechanisms.

Scale-locality of the cascade can be quantified by computing the energy transfer across scale $\ell$ that remains resolved when the velocity field is filtered at scale $0 \leq \ell^\prime \leq \ell$, here denoted $\Pi^{\ell,\ell^\prime}$.
That is, the energy cascade rate at $\ell$ is computed in the artificial absence of any motions smaller than scale $\ell^\prime$.
When $\ell^\prime = 0$, it is trivial that $100\%$ of the interscale energy transfer will be resolved, $\Pi^{\ell,0} = \Pi^\ell$. On the other hand, if $\ell^\prime = \ell$, then the resolved energy transfer is $\Pi^{\ell,\ell^\prime} = \Pi_{s1}^\ell + \Pi_{\omega 1}^\ell$. It is evident from Figure \ref{fig:Pi}(a) that this accounts for roughly $50\%$ of the total $\langle \Pi^\ell \rangle$. The question of scale-locality, then, is how quickly $\Pi^{\ell,\ell^\prime}$ approaches $\Pi^\ell$ when $\ell^\prime$ is decreased below $\ell$.

\begin{figure}
	\includegraphics[width=0.475\linewidth]{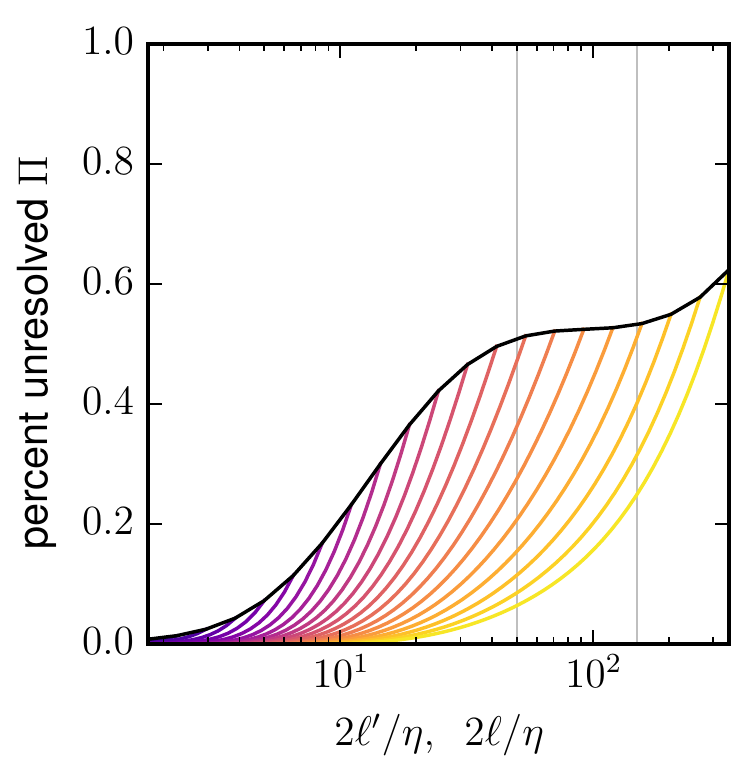}
	\includegraphics[width=0.51\linewidth]{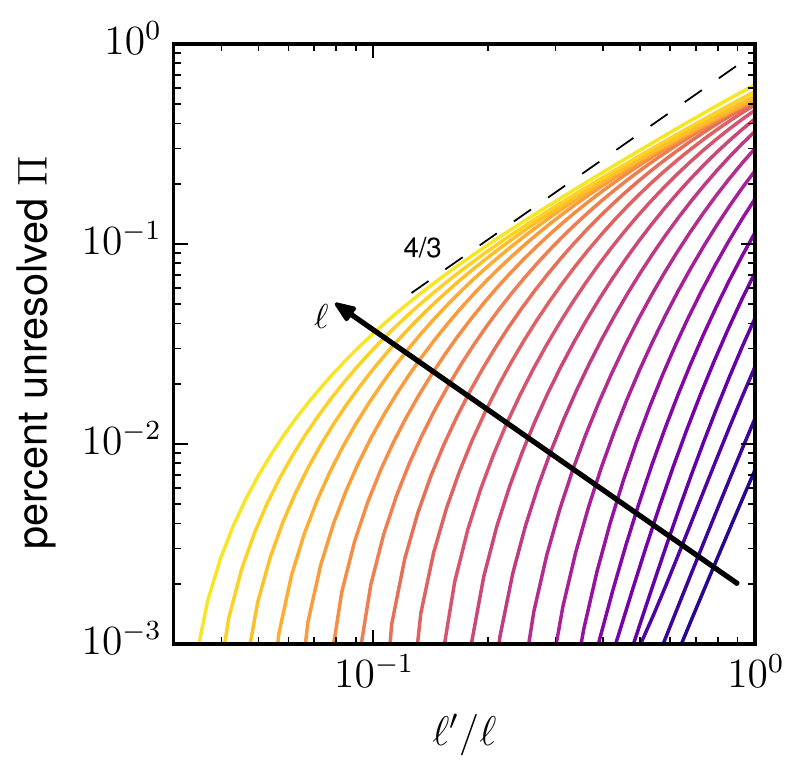}
	\caption{(a) Percent of $\langle \Pi^\ell \rangle$ unresolved when the velocity field is filtered at scale $\ell^\prime \leq \ell$, for various $\ell$. Top (black) curve indicates results for $\ell^\prime = \ell$. The (colored) curves emanating downward from the top curve show results for $\ell^\prime < \ell$, each curve representing a different fixed value of $\ell$. (b) The same results plotted log-log as a function of $\ell^\prime/\ell$, each curve again representing a different fixed value of $\ell$. The dashed line indicates the theoretical power-law exponent of $4/3$.}
	\label{fig:locality_all}
\end{figure}

First, a brief mathematical development is presented to generate the expressions needed to quantitatively evaluate the question of scale-locality outlined in words in the previous paragraph. Given Eq. \eqref{eq:moment-diffusion-solution}, the sub-filter stress tensor may be written as
\begin{equation}
\tau_\ell\left(u_i, u_j\right) = \int_{0}^{\ell^2} d\alpha~ \overline{\overline{A}_{ik}^{\sqrt{\alpha}} \overline{A}_{jk}^{\sqrt{\alpha}}}^\beta
\label{eq:subfilter-stress}
\end{equation}
where $\beta = \sqrt{\ell^2 - \alpha}$. Supposing the velocity field was known only to resolution $0 \leq \ell^\prime \leq \ell$, then the resolved amount of sub-filter stress would be
\begin{equation}
\tau_{\ell,\ell^\prime}(u_i, u_j)
= \int_{{\ell^\prime}^2}^{\ell^2} d\alpha~ \overline{\overline{A}_{ik}^{\sqrt{\alpha}} \overline{A}_{jk}^{\sqrt{\alpha}}}^\beta
+ \int_{0}^{{\ell^\prime}^2} d\alpha~ \overline{\overline{A}_{ik}^{\ell^\prime} \overline{A}_{jk}^{\ell^\prime}}^{\sqrt{\ell^2 - {\ell^\prime}^2}}
\label{eq:resolved-subfilter-stress}
\end{equation}
because for $\alpha < {\ell^\prime}^2$, the integrand is not fully known and the velocity gradients filtered at scale $\sqrt{\alpha}$ can at best be approximated by the velocity gradients filtered at scale $\ell^\prime$. Subtracting Eq. \eqref{eq:resolved-subfilter-stress} from Eq. \eqref{eq:subfilter-stress} results in an equation for the unresolved sub-filter stress,
\begin{equation}
\tau_{\ell}(u_i, u_j) - \tau_{\ell,\ell^\prime}(u_i, u_j)
= \int_{0}^{{\ell^\prime}^2} d\alpha~ \overline{
	\tau_{\beta^\prime}\left( \overline{A}_{ik}^{\sqrt{\alpha}}, \overline{A}_{jk}^{\sqrt{\alpha}} \right)	
}^{\sqrt{\ell^2 - {\ell^\prime}^2}},
\label{eq:unresolved-subfilter-stress}
\end{equation}
where $\beta^\prime = \sqrt{{\ell^\prime}^2 - \alpha}$. Contracting Eq. \eqref{eq:unresolved-subfilter-stress} with the filtered strain-rate tensor, $\overline{\mathbf{S}}^\ell$, defines the unresolved portion of $\Pi^\ell$ for a velocity field filtered at scale $\ell^\prime$. Figure \ref{fig:locality_all}a shows this quantity computed from DNS results. The top curve shows the result for $\ell^\prime = \ell$.  It is evident that the inter-scale transfer has at least some degree of locality, because these curves drop steeply from their originating points on the black $\ell^\prime = \ell$ curve.

Following \citet{Eyink2005}, a power-law relation may be expected,
\begin{equation}
\frac{\left\langle \Pi^\ell - \Pi^{\ell,\ell^\prime} \right\rangle}{\left\langle \Pi^\ell \right\rangle} \sim \left( \frac{\ell^\prime}{\ell} \right)^p.
\end{equation}
The exponent $p$ quantifies the extent of scale-locality. A steeper power-law means a higher degree of scale-local dominance and more adherence to cascade-like behavior. A log-log plot of the unresolved component of interscale energy transfer is shown in Figure \ref{fig:locality_all}(b). For large enough $\ell$ and $\ell^\prime$, the curves appear to collapse as would be a good indicator of scale-invariant behavior. Furthermore, the power-law curve fit produces an exponent near $1.1$, which deviates some from the theoretical prediction of $4/3$, though it is certainly steeper than a $2/3$ power-law which would result from perfect correlation across scales \citep{Eyink2005,Eyink2009}. This result suggests that current theories may over-estimate the de-correlation effect. A more precise estimate of this power-law requires higher resolution DNS.

\begin{figure}
	\includegraphics[width=0.49\linewidth]{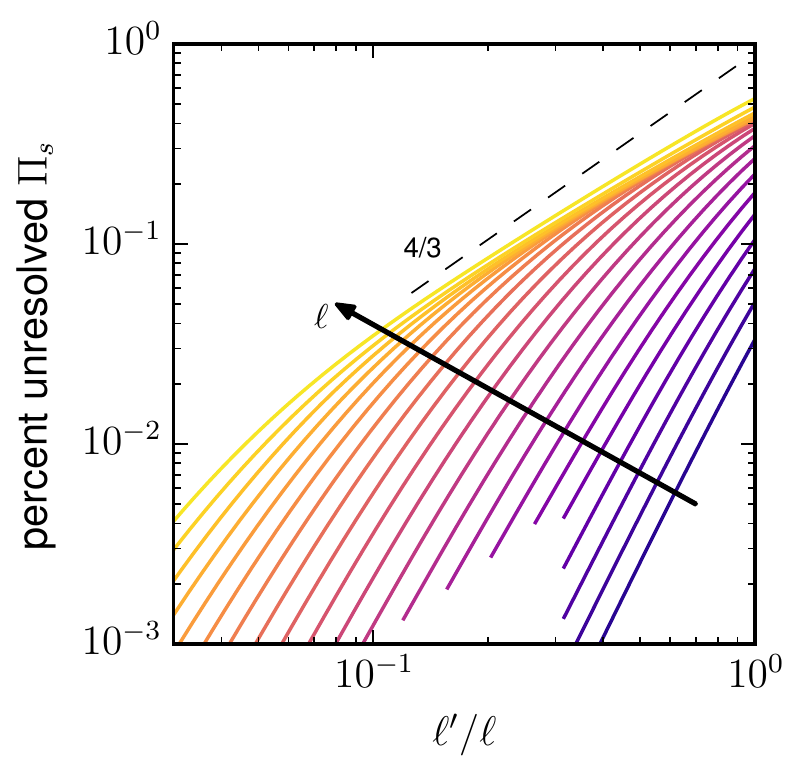}
	\includegraphics[width=0.49\linewidth]{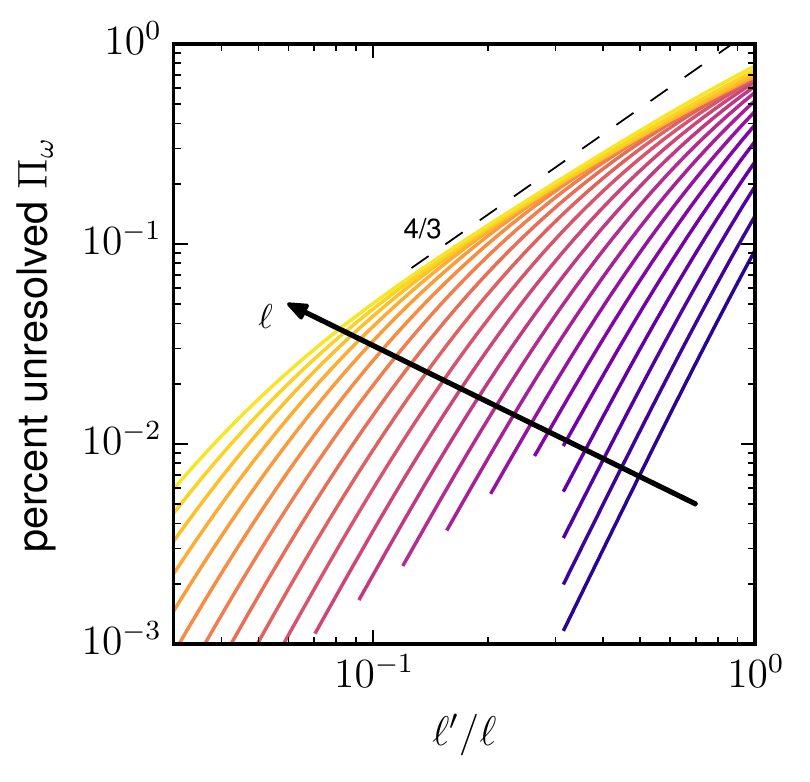}
	\caption{Percent of (a) $\langle \Pi_s^\ell \rangle = \langle \Pi_{s1}^\ell + \Pi_{s2}^\ell \rangle$ and (b) $\langle \Pi_\omega^\ell \rangle = \langle \Pi_{\omega 1}^\ell + \Pi_{\omega 2}^\ell \rangle$ unresolved by velocity field filtered at scale $\ell^\prime \leq \ell$ plotted log-log against $\ell^\prime/\ell$, as in Fig.\ \ref{fig:locality_all}(b). The dashed lines indicates a power-law exponent of $4/3$.}
	\label{fig:locality_components}
\end{figure}

The current formulation allows us to probe further with this analysis. Specifically, the above decomposition into resolved and unresolved components of the interscale energy transfer may be performed separately for strain self-amplification and vorticity stretching. Mathematically, this is done by replacing the full velocity gradient tensor in Eqs. \eqref{eq:subfilter-stress}-\eqref{eq:unresolved-subfilter-stress} with either its symmetric or anti-symmetric part. The results for strain self-amplification and vorticity stretching are qualitatively similar to Figure \ref{fig:locality_all}(a). In Figure \ref{fig:locality_components}, each of these two are shown on a log-log plot versus $\ell^\prime / \ell$. Similar collapse into power-law behavior is seen. This indicates that strain self-amplification and vorticity stretching have very similar scale-locality properties when it comes to their contributions to the energy cascade.

To further explore the observed discrepancy with theoretical predictions, the efficiency of the unresolved subfilter stress is computed as a function of $\ell^\prime$ for each filter width $\ell$. This decomposes the decay of the unresolved portion of $\Pi$ to be split into decreasing magnitude and de-correlation with $\mathbf{\overline{S}}^\ell$. The theory predicts a power-law decay of $2/3$ for each. To fully accomplish this, a slightly altered definition of efficiency is needed. In particular, to measure the de-correlation effect, the denominator of efficiency should use the trace of the unresolved subfilter stress tensor rather than the norm of its deviatoric part. Indeed, a perfectly decorrelated stress tensor in the physical sense could be an isotropic tensor. In other words, it is not desired to measure the extent to which the deviations from an isotropic tensor align with the filtered strain rate, but rather, how much energy transfer is occuring for a given level of subfilter activity. This is done by replacing the norm of the deviatoric part of the tensor in the denominator of Eq. \eqref{eq:efficiency} with the trace. It is then applied to Eq. \eqref{eq:unresolved-subfilter-stress}.

\begin{figure}
	\includegraphics[width=0.49\linewidth]{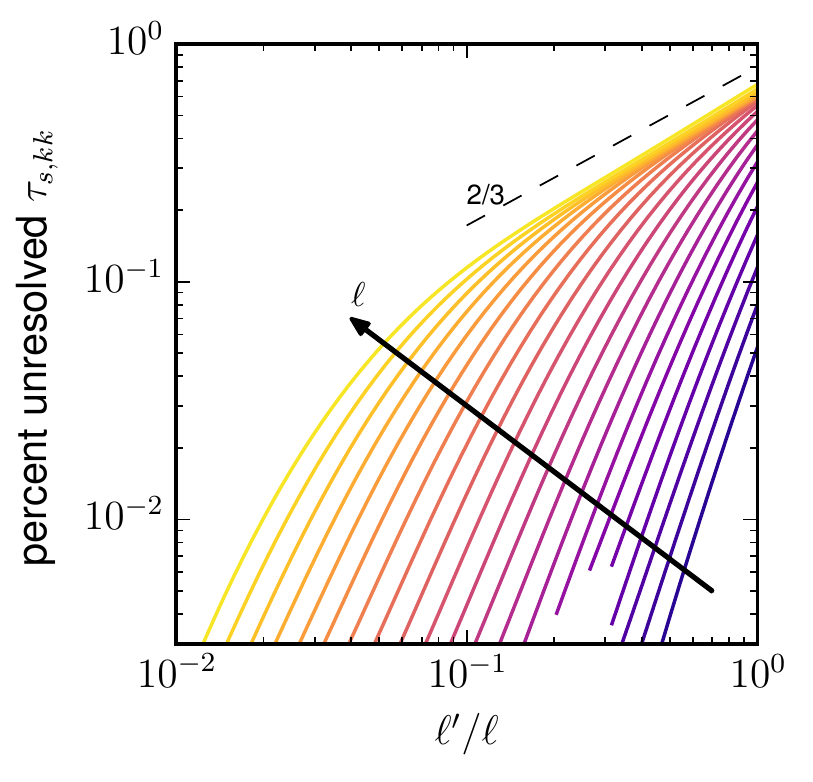}
	\includegraphics[width=0.49\linewidth]{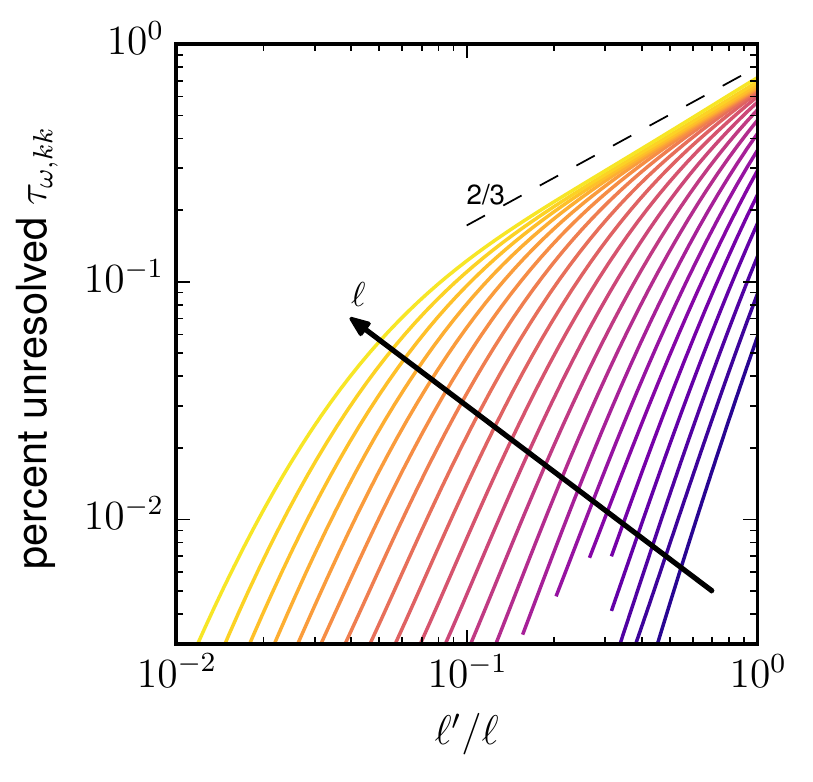}\\
	\includegraphics[width=0.49\linewidth]{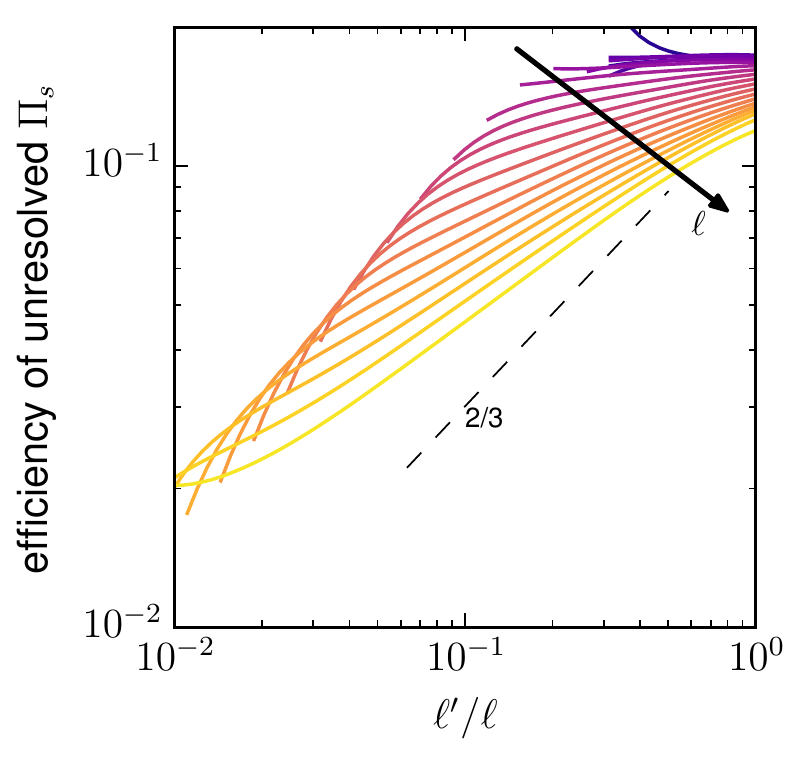}
	\includegraphics[width=0.49\linewidth]{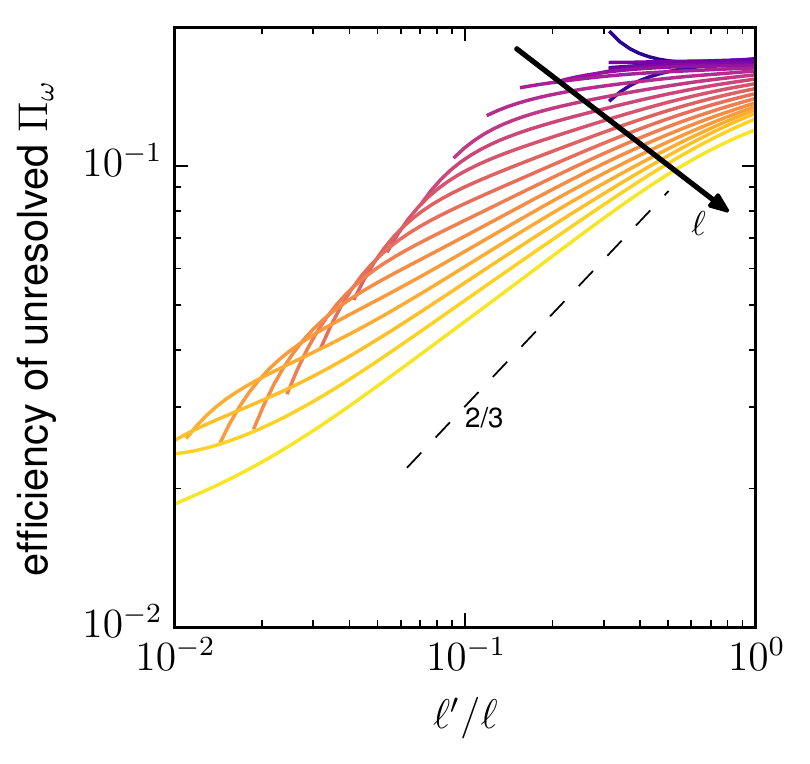}
	\caption{Trace of the subfilter stress tensor at scale $\ell$ remaining unresolved at scale $\ell^\prime$, as a function of $\ell^\prime / \ell$ for various values of $0.9\eta \leq \ell \leq 170\eta$: (a) unresolved strain-rate component, (b) unresolved vorticity component. Also, trace-based cascade efficiency of the unresolved subfilter stress: (c) unresolved strain-rate component, (d) unresolved vorticity component.}
	\label{fig:multiscale_efficiency}
\end{figure}

The results are shown in Figure \ref{fig:multiscale_efficiency}. The top left shows the decay of the unresolved trace for strain-rate amplification. Below it, the decay of efficiency is shown. The two right panels show the same results for vorticity stretching. For both mechanisms, the $2/3$ power-law is observed for the trace but the efficiency decays more slowly than predicted by theory. However, it the results suggest that the decorrelation (efficiency decay) becomes steeper as $\ell$ increases. It is plausible, then, that a simulation with a significantly higher Reynolds number may produce results in better agreement with theory at sufficiently large $\ell/\eta$. At the very least, the present results do no rule out that possibility.

\section{Conclusion}\label{sec:conclusion}

In this paper, the exact relationship introduced by \citet{Johnson2020} is exploited to discern the role of strain-rate amplification and vorticity stretching in transferring kinetic energy from large to small scales in turbulent flows. About half of the net energy cascade rate is proportional to velocity gradient production rates. These rates, which can be described as vorticity stretching and strain self-amplification of the filtered field, have important statistical biases which contain the signature of restricted Euler dynamics. For instance, the tendency toward a strain-rate eigenvalue state or $\lambda_1 = \lambda_2 = -\tfrac{1}{2}\lambda_3$ makes strain self-amplification more efficient at driving the cascade while depressing the efficiency of vorticity stretching. The efficiency PDFs indicate a tendency of turbulent dynamics to favor restricted Euler-like alignments. However, the pressure is also evidently important in determining average cascade rates. The pressure imposes the divergence-free condition, $\nabla\cdot\mathbf{u}$, one consequence of which is the three-to-one ratio between average strain-self amplification and vorticity stretching rates. Thus, incompressible Navier-Stokes dynamics tends to favor a higher energy cascade rate via strain self-amplification than vorticity stretching. This work joins \citet{Carbone2020} and \citet{Vela-Martin2020} in highlighting the crucial role of strain-rate self-amplification in the energy cascade, calling into question the classical emphasis on vorticity stretching \citep{Taylor1938,Onsager1949}.

Note that this conclusion about the relative contributions is for the average cascade rate and does not contradict the observation by others that vorticity stretching is more influential in generating extreme events in turbulent flows \citep{Buaria2019, Carbone2020}. Another point of nuance not yet pursued is a possible subtle difference between vortex stretching and vorticity stretching. Indeed, regions of strong vorticity are often positioned close to regions of strong strain \citep{Vlaykov2019}, so that the stretching of a vortex, if more broadly conceived of as the evolution of a flow structure that also includes regions of higher strain-rate, may include the strain self-amplification as defined mathematically in this paper. In general, further investigation of the \citet{Johnson2020} formulation in terms of spatially-coherent structures would be insightful \citep{Dong2020}.

A significant advantage of the formulation from \citet{Johnson2020} is the ability to write the remainder of the energy cascade rate in terms of multiscale vorticity stretching and strain-rate amplification mechanisms. It is shown in this paper that the stretching of smaller-scale vorticity and the amplification of smaller-scale strain-rate  by larger-scale strain-rate is responsible for the other half of the energy cascade rate. These subfilter stress components are more efficiently aligned with the filtered strain-rate tensor, proving to be better approximated using an eddy viscosity. The net contribution of multiscale interactions decreases with scale separation as a power-law, in close agreement to theoretical $4/3$ power-law predictions for the scale-locality of interscale energy transfer. This is true for both strain self-amplification and vorticity stretching mechanisms. Note that this is still not a particularly strong degree of locality, especially when compared with, e.g., the bubble breakup cascade in two-phase turbulent flows, where a much steeper power-law is predicted and observed \citep{Chan2020a, Chan2020b}.

Finally, a third cascade mechanism is explored, namely, the distortion of small-scale strain-vorticity covariance by the larger-scale strain-rate. While the net contribution of this term is negligible in the inertial range of three-dimensional turbulence, it has noticeable effects elsewhere. In two dimensions, it is the only possible mechanism, because both strain-rate amplification and vorticity stretching vanish exactly. This mechanism is responsible for the inverse cascade in 2D. It is argued, following \citet{Chen2006, Xiao2009}, that this term may be interpreted as a vortex thinning mechanism \citep{Kraichnan1976}. Furthermore, in three dimensions, the strain-vorticity correlation drives net backscatter opposing strain self-amplification and vorticity stretching for a range of filter width between viscous and inertial scales. This is the range of scales where the so-called spectral bump occurs, advancing vortex thinning as a candidate mechanism for what causes the bottleneck effect in 3D.

Velocity increments and structure functions have featured heavily in turbulence theory. No doubt, the ease of measurement using hot-wire anemometry (provided one may assume Taylor's hypothesis) has fueled the use of such two-point statistics. Filtered velocity gradients naturally encompass the information contained in velocity increments and similarly enable a scale-by-scale investigation of turbulent flows. The advantage of filtered velocity gradients is the ability to distinguish solid-body rotation from fluid deformation organized at scales much larger than $\eta$. Writing inter-scale energy transfer in terms of filtered velocity gradients not only connects with (and expands upon) classical cascade descriptions in terms of vorticity stretching, but also leverages the insights of the restricted Euler equation. A natural connection emerges between turbulence theory and modeling for large-eddy simulations, a tool of increasing practical importance in the age of supercomputing.

\section*{Acknowledgements}
This work was supported in part by the Advanced Simulation and Computing program of the U.S. Department of Energy’s National Nuclear Security Administration via the PSAAP-II Grant No. DE-NA0002373.
Ahmed Elnahhas is acknowledged for providing helpful feedback on an initial draft of this manuscript.
The author benefited from discussions on this topic with a number of colleagues, in alphabetical order by last name: Joseph Ballouz, Andy Bragg, Jose Cardesa, Greg Eyink, Sanjiva Lele, Adrian Lozano-Duran, Parviz Moin, Nick Ouellette, and Immanuvel Paul. 

\section*{Declaration of Interests}
The authors report no conflict of interest.

\appendix

\section{Relative importance of the small-scale strain-vorticity covariance}
\label{sec:app}

In Eq. \eqref{eq:Pi-decomposition}, the first two terms, $\mathring{\boldsymbol{\tau}}_{s1}$ and $\mathring{\boldsymbol{\tau}}_{\omega 1}$, are resolved at scale $\ell$. Therefore, in theory, these two need no approximate modeling for large-eddy simulations. The third and fourth terms, $\mathring{\boldsymbol{\tau}}_{s2}$ and $\mathring{\boldsymbol{\tau}}_{\omega 2}$, have been shown to be better approximated by an eddy viscosity, Figure \ref{fig:gama}. The fifth term, which may be thought of as a vortex thinning term, $\mathring{\boldsymbol{\tau}}_{c2}$, is not so easily modeled. The simplest model would be to ignore its contribution, which is close to zero in the mean within the inertial range of scales.

\begin{figure}
	\includegraphics[width=0.49\linewidth]{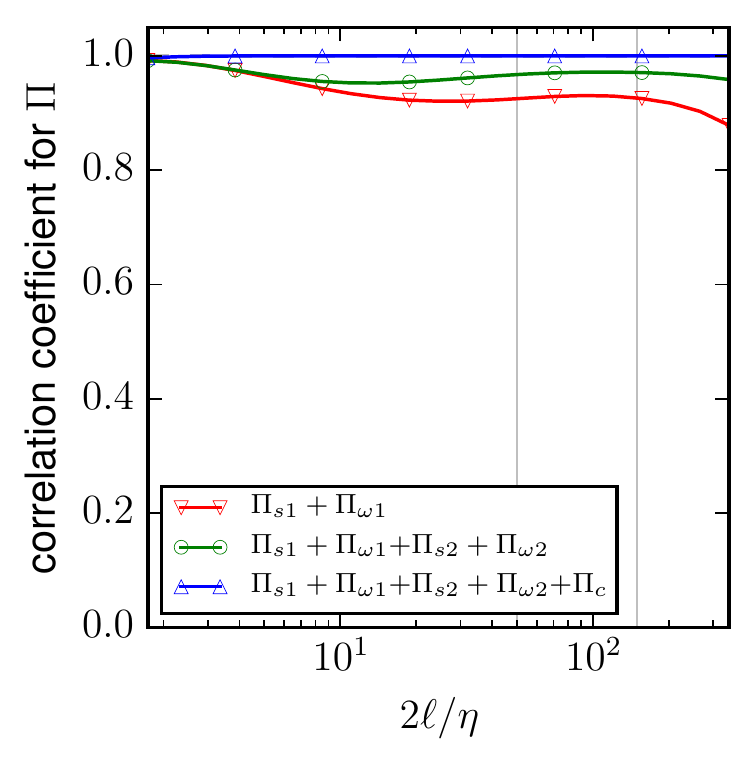}
	\includegraphics[width=0.495\linewidth]{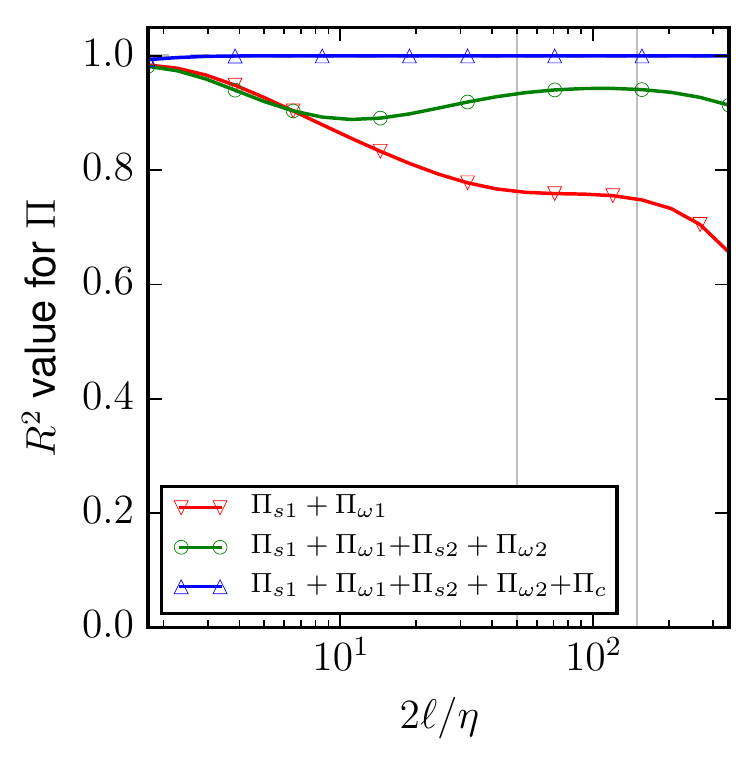}
	\caption{Correlation coefficient (a) and mean square error (b) for complete and incomplete sums of the cascade mechanisms.}
	\label{fig:accuracy}
\end{figure}

A preliminary \textit{a priori} study of the effects of neglecting some of the terms in Eq. \eqref{eq:Pi-decomposition} is shown in Figure \ref{fig:accuracy}. As previously shown by \citet{Borue1998}, the correlation coefficient of the tensor diffusivity model, $\Pi \approx \Pi_{s1} + \Pi_{\omega 1}$, is quite good, near $0.9$ in the inertial range. The normalized mean square error, $R^2$, is around 0.75.
When the multiscale vorticity stretching and strain self-amplification terms are added as well, $\Pi \approx \Pi_{s1} + \Pi_{\omega 1} + \Pi_{s2} + \Pi_{\omega 2}$, the correlation coefficient exceeds $0.95$ and the $R^2$ value also significantly improves. This observation provides evidence that the Eq. \eqref{eq:Pi-decomposition} is a meaningful decomposition of the energy cascade rate, because the accuracy of the sum improves as more terms are included. The correlation coefficients and $R^2$ values are perfect, $1.0$, when all five terms are included in the sum, reflecting the exact nature of Eq. \eqref{eq:Pi-decomposition} on a point-wise basis.

\bibliographystyle{jfm}
\bibliography{cascade}

\end{document}